\documentclass[sn-basic]{sn-jnl}%

\usepackage{makecell}
\usepackage{longtable}
\usepackage{multirow}
\usepackage{caption}
\usepackage{graphicx}
\usepackage{eurosym}
\usepackage[shortlabels]{enumitem}

\newcommand{\rev}[1]{\textcolor[rgb]{0.00,0.00,0.00}{#1}}

\jyear{2021}%

\theoremstyle{thmstyleone}%

\theoremstyle{thmstyletwo}%

\theoremstyle{thmstylethree}%

\begin{document}

\title[Negotiation Strategies in Ubiquitous Human-Computer Interaction]{Negotiation Strategies in Ubiquitous Human-Computer Interaction: A Novel Storyboards Scale \& Field Study}

\author*[1]{\fnm{Sofia} \sur{Yfantidou}}\email{syfantid@csd.auth.gr}

\author[2]{\fnm{Georgia} \sur{Yfantidou}}\email{gifantid@phyed.duth.gr}

\author[3]{\fnm{Panagiota} \sur{Balaska}}\email{pmpalask@phed.auth.gr}

\author[1]{\fnm{Athena} \sur{Vakali}}\email{avakali@csd.auth.gr}

\affil*[1]{\orgdiv{School of Informatics}, \orgname{Aristotle University of Thessaloniki}, \orgaddress{\city{Thessaloniki}, \postcode{54124}, \country{Greece}}}

\affil[2]{\orgdiv{Department of Physical Education and Sport Science}, \orgname{Democritus University of Thrace}, \orgaddress{\city{Komotini}, \postcode{69100}, \country{Greece}}}

\affil[3]{\orgdiv{Department of Physical Education and Sport Science}, \orgname{Aristotle University of Thessaloniki}, \orgaddress{\city{Thermi}, \postcode{57001}, \country{Greece}}}

\abstract{In today's connected society, self-tracking technologies (STTs), such as wearables and mobile fitness apps, empower humans to improve their health and well-being through ubiquitous physical activity monitoring, with several personal and societal benefits. Despite the advances in such technologies' hardware, low user engagement and decreased effectiveness limitations demand more informed and theoretically-founded Human-Computer Interaction designs. To address these challenges, we build upon the previously unexplored Leisure Constraints Negotiation Model and the Transtheoretical Model to systematically define and assess the effectiveness of STTs' features that acknowledge users' contextual constraints and establish human-negotiated STTs narratives. Specifically, we introduce and validate a human-centric scale, StoryWear, which exploits and explores eleven dimensions of negotiation strategies that humans utilize to overcome constraints regarding exercise participation, captured through an inclusive storyboards format. Based on our preliminary studies, StoryWear shows high reliability, rendering it suitable for future work in ubiquitous computing. Our results indicate that negotiation strategies vary in perceived effectiveness and have higher appeal for existing STTs' users, with self-motivation, commitment, and understanding of the negative impact of non-exercise placed at the top. Finally, we give actionable guidelines for real-world implementation and a commentary on the future of personalized training.}


\maketitle

\section{Introduction\label{introduction}}
 There is growing evidence that ubiquitous self-tracking technologies (STTs), such as wearables devices and mobile fitness applications, can be effective tools for motivating health behavior change, influencing people towards adopting a healthier and more active lifestyle \cite{orji2018persuasive,aldenaini2020trends,yfantidou2021self}. At the individual level, physically active people enjoy multiple health benefits, such as improved muscular and cardio-respiratory fitness, reduced symptoms of depression and anxiety, and lower coronary heart disease rates. Also, regular physical activity is proven to help prevent and manage non-communicable diseases such as heart disease, stroke, diabetes, and several cancers. At a collective level, more active societies can generate additional returns on environmental and social benefits, such as reduced use of fossil fuels, cleaner air, and healthier economies \cite{world2019global}. To seize these benefits, the World Health Organization recommends doing 75 to 150 minutes of moderate- and vigorous-intensity activity per week. However, one in four adults does not meet the recommended physical activity levels despite considerable personal and societal gains \cite{world2019global}. 
 
 To achieve desirable behavior change, more and more people are resorting to STTs, to reach their health and fitness goals. Wearables, for instance, have gained significant traction, with their global sales skyrocketing from less than 100 million in 2015 to an estimated 1.1 billion in 2022 \cite{vailshery_2021}. The concept behind such technologies is that through continuous monitoring, they can increase people's self-awareness of their everyday habits of physical activity, sleep, or mental health and empower them to alter their behaviors and pursue a healthier, more active life. Therefore, designing effective ubiquitous STTs has been a focus for many Human-Computer Interaction (HCI), and ubiquitous technology researchers \cite{adams2015mindless,tang2017harnessing,peters2018designing,niess2020exploring}. Research has shown that technology can be strategically designed to motivate desirable behavior change for better health and well-being. For instance, to help people adopt positive behaviors, such as regular exercise and healthy nutrition \cite{gouveia2015we,josekutty2017personalising,lee2011mining} or avoid negative or risky ones, such as prolonged sedentariness or even drug addiction \cite{wilde2018apps,mahmud2019wearables}. 
 
 Despite this growing interest and investment in designing effective STTs for mHealth with a focus on fitness and well-being, current technologies still suffer significant limitations: 
\begin{enumerate}[start=1,label={(\bfseries L\arabic*):}]
    \item \textbf{Lack of Theoretically-founded Design:} Most STTs adopt a non-theoretically-founded approach to their design \cite{yfantidou2021self,aldenaini2020trends}. Additionally, out of those that incorporate theoretical models, few acknowledge the existence of user contextual constraints, such as limited time or lack of confidence, that often lead to non-participation in the desired behavior \cite{yfantidou2021self,crawford1991hierarchical}. However, behavior change theories can positively affect STTs by informing design, guiding evaluation, and inspiring alternative experimental designs \cite{Hekler2013}. 
    \item \textbf{Dubious Feature Effectiveness:} STTs are characterized by the features they offer to their users, such as goal setting, social sharing, or virtual rewards. However, it is unclear which STTs' features are more effective in achieving successful health behavior change compared to others \cite{aldenaini2020trends}. On top of that, evaluating feature effectiveness usually entails lengthy and costly interventions and randomized control trials (RCTs) of small samples, and there are limited publicly available qualitative scales to be administered to larger populations in the initial development stages. Such uncertainty regarding STTs' feature effectiveness naturally often leads to unmet user expectations and eventually low user engagement and subsequent attrition, as reported in previous works \cite{attrition1,attrition2,clawson2015no}.
    \item \textbf{Limited Actionable Guidelines for STTs' Designers:} While several HCI works provide high-level insights for designing effective STTs \cite{epstein2020mapping,clawson2015no}, there is an absence of actionable guidelines that translate theory into practice. In other words, there is a gap between the abstract nature of behavioral theories and the concreteness of the design process. Hence, it is up to each designer to decide how exactly abstract theoretical concepts can be interpreted in terms of STTs' features, leading to increased time and effort investment or unfounded design decisions \cite{rabbi2020translating}. 
\end{enumerate}

Consequently, there is an increasing demand for STTs, especially those targeted at health behaviors, to be theoretically backed by behavioral change theories \cite{aldenaini2020trends,oyebode2021tailoring,orji2018persuasive}. Behavioral change theories attempt to interpret why human behaviors change, attributing such changes to environmental, personal, and behavioral factors. Recently, there has been increasing interest in the application of such theories in the health and technology domain with the hope that the understanding of behavioral change can improve offered services and quality of life \cite{neatu2015public,klasnja2017toward}. At the same time, there is substantial evidence that the use of theory in designing and implementing behavior change interventions can significantly improve their effectiveness \cite{national2007nice,michie2012theories}. \rev{Apart from informing design choices, behavior change theories can help guide evaluation strategies, inspire alternative experimental designs, evaluate qualitative data, and select target users \cite{hekler2013mind}}. However, as mentioned above, there is little knowledge about how these theories can be translated into STTs' features or which features might be more effective in achieving health behavior change.
 
To investigate how to translate behavioral change theories into concrete STTs' features and evaluate their perceived effectiveness, we conducted three preliminary studies of 248 participants. These studies examined how users respond to several features carefully designed to mirror elements from behavioral change theories. Specifically, we explored the Leisure Constraints Negotiation Model \cite{jackson1993negotiation}, as described in detail in Section \ref{relatd-work}, to account for constraints a user faces in the process of adopting a certain behavior, such as exercise participation and to make humans negotiators of their own self-tracking narrative. We designed an inclusive scale, StoryWear, that translates behavioral theory into technological practice and employed statistical analyses to ensure the reliability and validity of this scale and examine whether the usage of STTs plays a significant role in the process of behavior change for fitness and well-being.
Our results reveal that the newly-developed StoryWear scale demonstrates high reliability, rendering it a successful measurement model. Furthermore, our findings indicate a statistically significant difference in how users versus non-users of STTs approach health behavior change. Specifically, STTs' users showed higher commitment to physical activity and well-being and higher interest in technological features promoting behavioral change compared to non-users. It is worth mentioning that our results indicate that no other independent variable, such as user gender, educational level, or age, plays a statistically significant role in the users' preferences toward health behavior change through wearables. Our work offers three main contributions in the domain of ubiquitous HCI design:
\begin{enumerate}[start=1,label={(\bfseries C\arabic*):}]
    \item \textbf{Users as Negotiators of Constraints:} We utilize and showcase for the first time, to the best of the authors' knowledge, the importance of the Leisure Constraints Negotiation Model for the design and development of STTs' features. This model accounts for the constraints users face in their daily lives, such as lack of time, money, self-confidence, or social encouragement, that inhibit or prohibit participation and enjoyment in physical activity and well-being. By adopting such a theoretical model, we change the role of users from passive viewers of constraints to active negotiators through them while enabling the design of theoretically-founded STTs (L1).
    \item \textbf{The Inclusive StoryWear Scale:} We develop and make publicly available an inclusive by-design scale, called StoryWear, for theoretically-backed HCI in self-tracking. Due to its novel storyboards format, our scale can be administered to a broader participant pool, including people of low technological literacy, and young or old age, enabling more inclusive and diverse mHealth interventions. We validate our scale through three distinct field studies and compare the perceived effectiveness of several STTs' features. We show that negotiation strategies and their respective features differ in their effectiveness in motivating health behavior change in physical activity and well-being (L2).
    \item \textbf{Actionable Guidelines for STTs' Designers:} We demonstrate that the usage of STTs can positively contribute to the behavior change journey. Based on our findings, we offer concrete design guidelines and STTs' feature suggestions that appeal to a broad audience based on the Leisure Constraints Negotiation Model. Such guidelines can help HCI and ubiquitous computing designers and practitioners focusing on self-tracking to incorporate behavioral change theory into their design for creating successful health behavior change interventions (L3).
\end{enumerate}

We structure the rest of this paper as follows: In Section~\ref{relatd-work}, we give an overview of the related work in the fields of Behavioral Change Theories and Ubiquitous HCI Design for mHealth. In Section~\ref{study}, we present our study's design process, scale, and methodology, and in Section~\ref{results}, we showcase our work's most important results. In Section~\ref{discussion}, we discuss how our findings can be applied in designing STTs that exploit behavioral change theories, and in Section~\ref{uptake}, we explore the practical and theoretical implications of our work. Finally, in Section~\ref{conclusions}, we conclude our work, comment on its limitations, and provide directions for future work.
 
\section{Related Work} \label{relatd-work}
In this section, we present an overview of behavioral change theories (subsection~\ref{theories}) and mHealth ubiquitous interventions employing such theories in their design (subsection~\ref{theories-design}). To conclude, we identify research gaps in current literature and compare previous work with the contributions of our approach. 

\subsection{Theories of Behavior Change}\label{theories}
A deep understanding of health behaviors and their context is the foundation of most successful public health programs and initiatives \cite{glanz2008health}. Therefore, mHealth interventions that aim at improving the users' health and well-being can be best designed given an understanding of the underlying theories of behavior change and an ability to utilize them successfully. In recent years, there has been growing interest in theories and models of behavior change drawn from psychology, economics, and sociology \cite{michie2008theory}. These theories include but are not limited to Bandura's Social Cognitive Theory \cite{bandura1999social}, Ajzen's Theory of Planned Behaviour \cite{ajzen1991theory}, or, more recently, Fogg's Behavior Model for persuasive design \cite{fogg2009behavior}. Such theories describe how behaviors develop and change over time as a factor of psychological, social, and contextual factors such as habits and routines \cite{darnton2008practical}. 

Amongst the most commonly applied theories of behavior change, \cite{eufic_2014} is Prochaska's Transtheoretical Model, also referred to as the "Stages of Change" model \cite{prochaska1997transtheoretical}. The Transtheoretical Model introduces five stages of behavior change: pre-contemplation, contemplation, preparation for action, action, and maintenance. During the pre-contemplation stage, an individual, with or without awareness of the problem, has no thought of changing their behavior. Transitioning to the contemplation stage, one starts thinking about changing a specific behavior. In the preparation stage, the individual creates a change plan, while in the action stage, one starts exhibiting the newly formed behavior consistently. Finally, an individual enters the maintenance stage once the behavior has become habitual, performed consistently for six or more months. An issue with the stages of change model is that it is possible for a person to relapse, even after entering the maintenance stage, and fall back into earlier stages \cite{MIMIAGA2009203}. Factors, also known as constraints, that can contribute to such relapse include external weather conditions or seasonal changes and internal factors, such as personal issues.

For years, behavior change and leisure constraints research had supported the conception that constraints are insurmountable obstacles to behavior adoption. In other words, it has been assumed that if an individual faced a constraint, the outcome would be equal to non-participation in the desired activity required for change, such as exercise participation or smoking cessation. Subsequently, that would mean that ubiquitous HCI designers who have no control over such external factors would be incapable of designing STTs that enable users to overcome such constraints. However, the Leisure Constraints Negotiation Model \cite{jackson1993negotiation} supports that leisure participation "is dependent not on the absence of constraints but on negotiation through them". In other words, the proposed concept is that individuals are not bound by their constraints but can apply different negotiation strategies to strike a balance between constraints and motivations. Hence, integrating such negotiation strategies into STTs' design would enable HCI designers to empower their users to overcome the constraints presented upon them, such as lack of time, money, confidence, or social support.

The fundamental assumption of negotiation is based on social cognitive theory. According to this theory, individuals are likely to actively choose or change situations and environmental conditions that are known to influence their behavior rather than passively accepting unfavorable situations \cite{maddux1993social}. In other words, the basic assumptions of negotiation are derived from ``compatible rather than competitive'' perspectives, which view people as ``active shapers'' rather than ``passive reactors'' \cite{mannell2005don}. People can and often do participate in leisure activities despite the obvious barriers. Studies have begun to show that some barriers to participation can be overcome and that through negotiation, constraints actively shape individual leisure expression by interacting with preferences and behavioral patterns \cite{jackson1995negotiation,nadirova2000alternative,jackson1993negotiation}.

In this work, we utilize a high-validity scale \cite{balaska2019} that draws from the Transtheoretical Model and the Leisure Constraints Negotiation Model and builds upon and modernizes previous scales on the field \cite{marcus1992stages,alexandris2007investigating}. We utilize such a scale because it is domain-specific to physical activity and well-being and is more in line with recent technological advancements than older scales, meaning it is more straightforward to translate its dimensions into STTs' features. Additionally, it has been validated with Greek sample populations. The scale identifies eleven negotiation dimensions consisting of 3 negotiation strategies each, defined in Table~\ref{tab:dimensions}, namely Knowledge, Negative Impact of Non-Exercise Understanding, Self-motivation, Enable, Socialization, Enhancement, Commitment, Creating Pulse, Encouragement Relations Development, Time Management, and Financial Management. 
\begin{table}[htb!]
\centering
\captionsetup{justification=centering}
\caption{Definitions of the eleven dimensions of negotiation strategies for exercise participation.\label{tab:dimensions}}
\resizebox{\textwidth}{!}{%
\begin{tabular}{|l|l|}
\hline
\rowcolor[HTML]{9B9B9B} 
\multicolumn{1}{|c|}{\cellcolor[HTML]{9B9B9B}{\color[HTML]{FFFFFF} \textbf{Negotiation Dimension}}} & \multicolumn{1}{c|}{\cellcolor[HTML]{9B9B9B}{\color[HTML]{FFFFFF} \textbf{Explanation}}} \\ \hline
\rowcolor[HTML]{EFEFEF} 
Knowledge & Strategies to collect new information and raise awareness about exercise \\ \hline
Negative Impact of Non-Exercise Understanding & Strategies to exploit emotional reactions associated with the negative effects of non-exercise \\ \hline
\rowcolor[HTML]{EFEFEF} 
Self-motivation & Strategies to reassess personal values for exercise participation \\ \hline
Enable & Strategies to use exercise as a substitute behavior to tackle everyday slack \\ \hline
\rowcolor[HTML]{EFEFEF} 
Socialization & Strategies for finding exercise buddies and increased socialization \\ \hline
Enhancement & Strategies to utilize self-motivation means to achieve strategic objectives in exercise \\ \hline
\rowcolor[HTML]{EFEFEF} 
Commitment & Strategies of personal commitment growth for practicing \\ \hline
Creating Pulse & Strategies for creating external stimuli to increase motivation for exercise participation \\ \hline
\rowcolor[HTML]{EFEFEF} 
Encouragement Relations Development &  Strategies for the acceptance and support from the social environment for exercise participation \\ \hline
Time Management & Strategies to include exercise in the daily individual program and save time for exercise participation \\ \hline
\rowcolor[HTML]{EFEFEF} 
Financial Management & Strategies for saving money needed for exercise participation \\ \hline
\end{tabular}%
}
\end{table}

\subsection{Theories of Behavior Change in HCI Design}\label{theories-design}
Various works have proposed incorporating psychological, behavioral, and behavioral economics theories in the HCI design of ubiquitous computing, offering considerable contributions in the field of theoretically founded design.

Early research works in the field of ubiquitous technology for health behavior change include Lin et al.'s Fish'n'Steps (2006) \cite{lin2006fish}, Consolvo et al.'s Houston (2006) \cite{consolvo2006design}, and UbiFit Garden (2008) \cite{Consolvo2008,consolvo2009theory}. These works utilize the Transtheoretical Model, the Goal-Setting Theory, and the Cognitive Dissonance Theory, among others, to design and implement RCTs for evaluating STTs' feature effectiveness in improving users' physical activity and well-being. All three report positive results regarding theoretically-backed health behavior change interventions. Similarly, \citet{munson2012exploring} explore goal-setting, rewards, self-monitoring, and sharing to motivate physical activity through the development and evaluation of two mobile applications, GoalPost and GoalLine. Their results highlight the benefits of both secondary and primary goals and nonjudgmental reminders. Additionally, they reveal motivational shortcomings with commonly used features, such as trophies and ribbons, raising questions with regard to effective feature design. 

Inspired by Behavioral Economics research, \citet{lee2011mining} experiment with the presentation and timing of choices to encourage people to make self-beneficial decisions through self-tracking. Specifically, they apply three persuasion techniques, the default option strategy, the planning strategy, and the asymmetric choice strategy, to promote
healthy snacking in the workplace, showcasing positive results with regard to the exploitation of theoretically-founded design in STTs for promoting user well-being. While their contributions are significant in the field of ubiquitous technology research, these studies suffer from limitations deriving from the nature of RCT-based evaluation: they have high requirements in time and money, are usually smaller-scale, and are difficult to reproduce. On the contrary, our work adopts a different approach to evaluation, i.e., survey-based, which is more suitable for the early stages of technology development due to its rapid nature, reproducibility across different settings, and low deployment cost. Additionally, while these interventions make use of the Transtheoretical Model, a combination of the Transtheoretical Model and the Leisure Constraints Negotiation Model has not been utilized before for the HCI design of STTs.

On a different note, numerous studies provide HCI researchers and practitioners with insights or actionable guidelines for designing systems based on behavioral and psychological theories. \citet{neatu2015public} explores how behavioral economics principles, such as nudging \cite{leonard2008richard}, availability bias \cite{tversky1974judgment}, and illusion of control \cite{langer1975illusion}, can be applied for developing effective public health strategies that target behavior change. While the authors encourage the interdisciplinarity between behavioral economics and wearable design, they do not provide any guidelines for HCI design. 

To bridge this gap, \citet{peters2018designing} and \citet{rapp2019design} propose concrete guidelines for HCI design based on psychological theories concerning how technology designs support or undermine basic psychological needs. Specifically, Peters et al. introduce the Motivation, Engagement, and Thriving in User Experience (METUX) model to  explain psychological needs in the context of HCI. METUX is based on psychological research, casting light on how technology designs support or undermine basic user needs. Similarly, Rapp et al. analyze four design fictions to explore the diverse social and psychological long-term consequences of health behavior change technologies. However, none of the two studies evaluate the proposed guidelines' perceived effectiveness through a field study, nor do they focus on emerging STTs, such as wearables. To this end, \citet{mejova2019effect} conduct a large-scale user study including data collection. Based on psychometric and demographic variables, as well as browsing and application use data analysis, they draw conclusions and provide design suggestions about employing this knowledge in the design, personalization, and deployment of health behavior change technologies. Their suggestions include tracking of a single activity, incorporation of the user's socioeconomic status, integration of the user's values in the interaction or interface, and consideration of application usage in predictive analytics. While these recommendations are based on real-world user data, their effectiveness remains untested, and they are also not formalized into a concrete framework. \rev{Similarly, \citet{li2011understanding}, through a series of user interviews, identify six types of questions that users ask regarding personal informatics data, i.e., status, history, goals, discrepancies, context and factors, and two phases of reflection, i.e., discovery and maintenance, and provide concrete feature guidelines for each phase. However, their interview-based evaluation does not allow for testing the reproducibility of their results under different settings, contrary to our survey-based approach.} Our work translates theoretical concepts drawn from the Transtheoretical Model and the Leisure Constraints Negotiation Model into STTs' features, and we evaluate their effectiveness through three field studies. Based on the studies' results, we propose a set of concrete guidelines for effective, theoretically founded HCI design for ubiquitous STTs.

Similar to our work, a number of works have conducted field studies to evaluate the effectiveness of their HCI design guidelines. For instance, \citet{adams2015mindless} propose the concept of "Mindless computing" based on dual-process theories \cite{cacioppo1984elaboration,todorov2002heuristic}, and the theory of "nudging" \cite{leonard2008richard}. They also evaluate its potential and provide design considerations based on three small-scale pilot studies. However, their design guidelines do not necessarily refer to ubiquitous STTs, rather than HCI as a whole. 

\citet{gouveia2015we} focus on the design of ubiquitous wearable technologies by exploring the design space of glanceable feedback for physical activity trackers. While they conduct a field study to evaluate the effectiveness of their designs, they do not specify the theoretical foundation they are based on, contrary to our approach. In related work, \citet{duro2019motivational} study the motives of regular and non-regular exercisers’ user engagement with motivational text messages and the importance of visual presentation. Using the self-determination theory \cite{deci2012self}, they provide preliminary knowledge regarding motivational messages acquired through user surveys. Research on an individual's intrinsic motivation in recreation and exercise has shown a large effect on participants' behavior \cite{palen2011mixed,weissinger1992relation}. External motivation in the same domain is related to the existence of external rewards, such as social recognition and positive criticism \cite{deci2013intrinsic}, and is more pronounced in competitive professional sports and recreational sports, such as amateur sports or leisure time exercise \cite{alexandris2002perceived,alexandris2007investigating}. An example of external motivation in recreational activities is good health, which is associated with good appearance and social recognition \cite{rotter1966generalized,ryan2008self,white1959motivation}. However, prior work focuses only on one design element (motivational messages), and it does not propose any guidelines for developing and structuring motivational messages. 

Finally, utilizing dual-process theories of decision making \cite{strack2004reflective}, the theory of "nudging" \cite{leonard2008richard}, and Fogg's Behavior Model \cite{fogg2009behavior}, \citet{caraban2020nudge} present the design and evaluation of the "Nudge Deck", a card-based, design support tool for the design of technology-mediated nudges. While the tool is an excellent resource for assisting HCI design, it assumes equal effectiveness of all nudging features across the population. On the contrary, in our work, we evaluate the effectiveness of each STTs' feature based on our field studies' results. 

\rev{To cover the identified gaps in the literature and overcome the limitations of RCT-based evaluation, prior research has introduced some measurement instruments in the domain of HCI. For example, the Technology-Supported Reflection Inventory evaluates if and how an interactive technology helps a user reflect \cite{bentvelzen2021development}, while the popular User Burden Scale evaluates the difficulty of use \cite{suh2016developing}. Nevertheless, our contribution is complementary if not orthogonal to these or similar scales in the domain since we ``measure'' a distinct concept, namely the effectiveness of negotiation in health behavior change through self-tracking technology. Inspired by these works, we designed a novel, inclusive storyboards scale and evaluated it through different field studies as described in Section \ref{study}. To the best of the authors' knowledge, this is the first scale to use a combination of the Transtheoretical Model and the Leisure Constraints Negotiation Model for the HCI design of ubiquitous computing. }

\section{Study Design and Methodology} \label{study}
\begin{figure}[htb!]
  \centering
  \includegraphics[width=.9\linewidth]{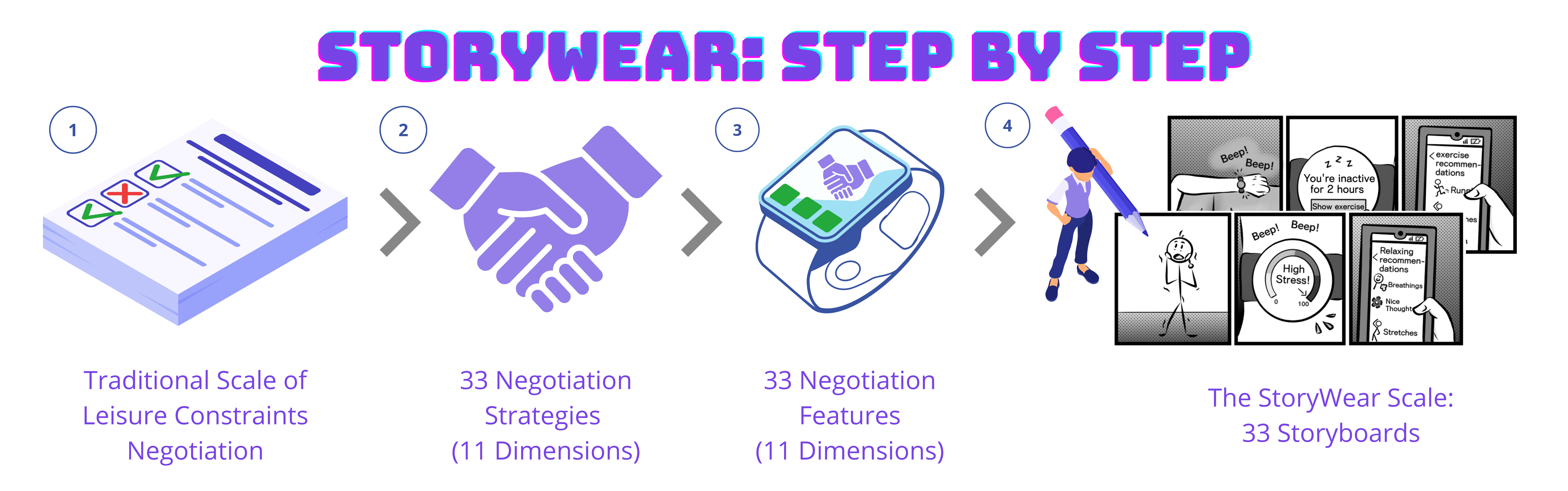}
  \caption{An overview of StoryWear's design process.\label{fig:methodology}}
\end{figure}

We designed our study to investigate the relationship between STTs and negotiation strategies (knowledge, understanding of the negative impact of non-exercise, self-motivation, enable, enhancement, commitment, creating pulse, socialization, time management, financial management \cite{balaska2019}) as a means of health behavior change, motivating a more active and healthier lifestyle. Figure \ref{fig:methodology} shows an overview of the design process.

\noindent\textbf{{The StoryWear Scale.}} To collect data to investigate our hypothesis, we introduce a novel storyboards-based scale, the so-called ``StoryWear: A \textbf{Story}boards Scale for Negotiation Strategies in \textbf{Wear}able Computing''. First, we start from a traditional, textual scale described in our earlier work \cite{balaska2019} (Step 1, Figure \ref{fig:methodology}). This scale contains eleven negotiation dimensions and 33 items, referred to as negotiation strategies, as described in Section~\ref{theories} (Step 2, Figure \ref{fig:methodology}). Notice that the items within each dimension do not relate to technology in the original scale but only reflect the actions individuals take to overcome obstacles preventing them from performing the desired behavior. Through an iterative refinement process, we initially translate these 33 negotiation strategies into an equivalent number of ubiquitous STTs' features, so-called "negotiation features" (Step 3, Figure \ref{fig:methodology}). For the item-to-item mapping, please refer to Appendix~\ref{ap:features}. 

Specifically, in the \textit{``Knowledge''} dimension, we include features that enable users to learn more about exercise through STT, such as notifications about new books and articles with regard to physical activity and its benefits or promotion of fitness influencers' content. Regarding the \textit{``Negative Impact of Non-Exercise Understanding''}, our negotiation features range from simple information provision to personalized dramatic portrayals of the effects of non-exercise on the individual's health. In the \textit{``Self-motivation''} dimension, we introduce real-time and offline cause-and-effect linkage features that highlight the short-term and long-term effects of exercise on the person's health, welling, and appearance, in an effort to motivate the users to continue their effort. The \textit{``Enable''} dimension contains features that encourage easy-to-follow or relaxing forms of physical activities that can stir the user in the right direction in their fitness journey. The negotiation features of the \textit{``Socialization''} dimension incorporate the social aspect of fitness and well-being by encouraging the participation of the users' social circle in exercising. The \textit{``Enhancement''} dimension includes features that offer material and virtual rewards to users, as well as opportunities to achieve them through maintaining and enhancing their exercise routine. Concerning \textit{``Commitment''}, we introduce negotiation features that facilitate the user to commit to their exercise plan through realistic goal-setting, personal and group challenges, and social sharing. When it comes to the \textit{``Create Pulse''} dimension, features include just-in-time reminders for picking up sports equipment or notifications about new sports clothing and accessories relevant to the user's favorite fitness activity. In the \textit{``Encouragement Relations Development''}, negotiation features promote contact with and support from real and virtual coaches, the user's social circle, as well as users with similar interests. Regarding the \textit{``Time''} dimension, features are oriented towards optimizing the user's exercise planning by utilizing just-in-time reminders, contextualizing it to the user's everyday routine, or adapting it to the user's schedule. Finally, in the \textit{``Financial''} dimension, features are aimed at increasing awareness of the cost of non-healthy life choices or budgeting and information provision for covering exercise costs.

Then, we present each negotiation feature above in a storyboard depicting a ubiquitous STT for promoting a healthier lifestyle in terms of physical activity and general well-being (Step 4, Figure \ref{fig:methodology}). A storyboard is "a graphic organizer that consists of illustrations or images displayed in sequence for the purpose of pre-visualising a motion picture, animation, motion graphic or interactive media sequence" \cite{wikipedia_2021}. Depicting negotiation strategies in storyboards facilitates eliciting responses from diverse populations since storyboards provide a common visual language that individuals can understand more easily compared to traditional means, regardless of their technological literacy, educational level, or age, to mention a few \cite{van2006value}. Especially when it comes to emerging technologies, such as wearables, which the general population might not be fully familiarized with, a visual depiction can significantly enhance understanding of complex technological notions. At the same time, storyboards have been proven successful in depicting strategies based on prior work \cite{orji2017towards,busch2016more,chen2019understanding}. 

To develop the initial storyboard drafts, we follow the design guidelines of \citet{10.1145/1142405.1142410}. Specifically, each collaborator designs their storyboard sequences individually to decrease the effect of peer-opinion influence. Then researchers gather, compare designs, and conclude with the most appropriate depiction for each negotiation feature. Finally, according to the same design protocol, an artist was recruited to design the 33 final storyboards. We evaluated and iteratively refined these storyboards following expert discussions with both Sports Science and Computer Science researchers, as well as STTs' users without an academic background, which led to some minor refinements. All storyboards depict a gender-neutral stick figure and its interaction with a ubiquitous mHealth system or application for promoting a healthier, more active lifestyle. Based on Truong's guidelines, we utilize some text to facilitate the understanding of novel applications and features. Also, we include people-like figures in the storyboards to elicit responses related to the interaction experience, and we explicitly indicate the passage of time only when time is relevant. Finally, we use the minimal level of detail required to highlight the prominent features of the system. 

\begin{figure}[htb!]
  \centering
  \includegraphics[width=.7\linewidth]{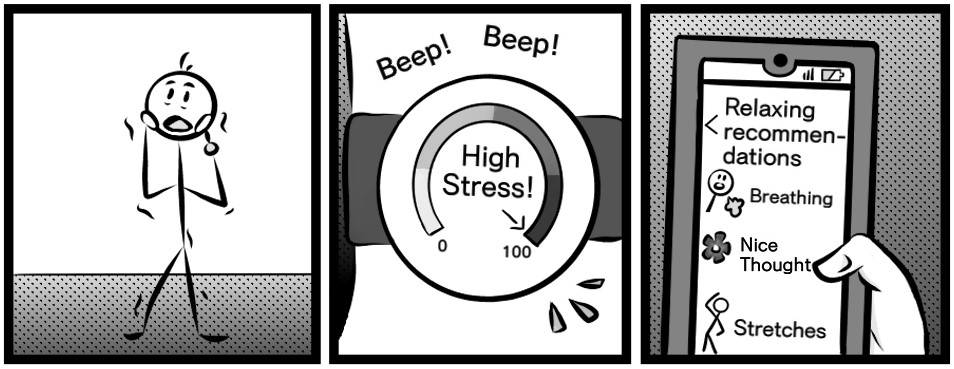}
  \caption{Dimension Enable - 3$^{rd}$ item: A feature for stress awareness and management.\label{fig:example}}
\end{figure}
\begin{figure}[htb!]
  \centering
  \includegraphics[width=.7\linewidth]{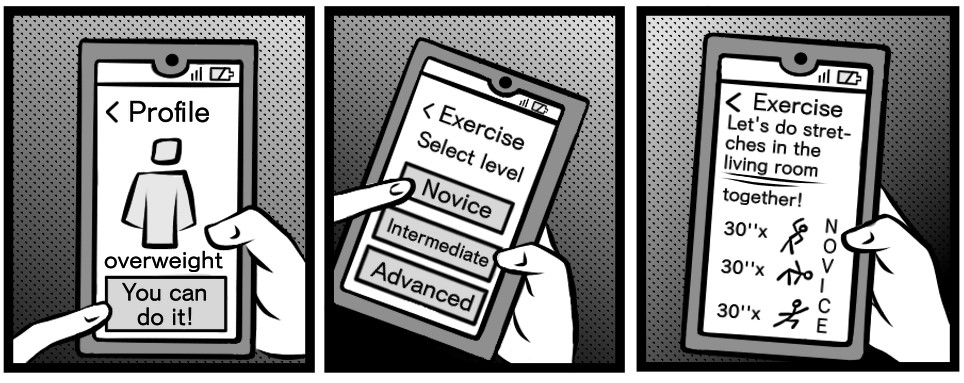}
  \caption{Dimension Commitment - 1$^{st}$ item: A feature for realistic goal-setting based on user traits.\label{fig:example2}}
\end{figure}
Figure~\ref{fig:example} shows an example storyboard from our scale depicting an item from the "Enable" dimension of negotiation strategies, i.e., "I see exercise as a stress relief mechanism when I feel tense". The researchers textually translate the original item into a stress awareness and management feature, which is then visually depicted through the presented storyboard. Similarly, Figure~\ref{fig:example2} presents a storyboard depicting an item from the "Commitment" dimension, i.e., "I persuade myself that I am capable of exercise". The original item is then translated into a realistic, user traits-based goal-setting feature and visually depicted through the storyboard. 

To elicit feedback about the effectiveness of different negotiation features through storyboards, each item in the StoryWear scale is followed by the question "How often would you use this feature?" and a 7-point Likert scale ranging from "1-Never" to "7-Always", in accordance to the original scale \cite{balaska2019}. We also include questions for collecting participants' demographic information and details about their familiarity with and usage of STTs.

The full StoryWear scale, consisting of 33 storyboards matching the 33 items of the original negotiation strategies scale, can be found in the supplementary material and used with attribution for further studies.
\bigbreak
\noindent\textbf{Sample and Process.} We recruited participants for three different preliminary studies to evaluate the reliability and validity of the StoryWear scale and test our hypotheses with diverse sample populations. The first sample consists of 104 recreational athletes exercising in indoor and outdoor gyms in Athens, the capital of Greece. The second sample consists of 40 professional swimmers training in Athens or Komotini, a university city in northern Greece. Finally, the third sample consists of 101 everyday people who follow urban exercise routes within Thessaloniki, the second-largest city in Greece, jogging, walking, or using outdoor exercise equipment. 

Note that we conducted the field studies amidst the Covid-19 pandemic in compliance with the national and regional restrictions. Specifically, gyms throughout Greece were closed during the field study; hence we reached out to past gym-goers through online communication. Similarly, professional athletes were contacted via e-mail to complete the scale. However, everyday people following urban exercise routes were contacted in person during their activity, taking all necessary precautions to protect public health.

\section{Results} \label{results}
The main goal of this paper is to examine the perceived effectiveness of the Leisure Constraints Negotiation Model and its negotiation dimensions interpreted as ubiquitous STTs' features. To achieve this, we have utilized various commonly used analytical tools and procedures, as this section summarizes. Initially, we provide details about sample demographics and STT usage. Then, we present the internal consistency reliability assessment for the StoryWear scale, followed by our results concerning the perceived effectiveness of negotiation features. 

\medskip
\noindent\textbf{Participants' Demographics.} In total, our full sample consists of 248 participants, having filtered out incomplete responses. The participants received no monetary compensation for their participation in the study. All participants were at least 17 years old at the time of data collection. In general, we have achieved recruiting a diverse population in terms of gender, age, educational level, occupation, income, and marital status as seen in Table~\ref{tab:demographics}.

\begin{table}[htb!]
\centering
\captionsetup{justification=centering}
\caption{Participants' demographic information.\label{tab:demographics}}
\resizebox{.6\textwidth}{!}{%
\begin{tabular}{llll}
\cline{1-4} 
\multicolumn{2}{l}{} & Frequency & Percentage \\ \cline{1-4} 
\multirow{2}{*}{Gender} & Men & 109 & 44.5\% \\ \cline{2-4} 
 & Women & 134 & 54.7\% \\ \cline{1-4} 
\multirow{2}{*}{Age} & 17-39 & 163 & 66.5\% \\ \cline{2-4} 
 & 40-59 & 82 & 33.5\% \\ \cline{1-4} 
\multirow{4}{*}{Education} & High School graduate & 84 & 34.3\% \\ \cline{2-4} 
 & Vocational School graduate & 40 & 16.3\% \\ \cline{2-4} 
 & University graduate & 91 & 37.1\% \\ \cline{2-4} 
 & Post-graduate & 30 & 12.2\% \\ \cline{1-4} 
\multirow{8}{*}{Occupation} & University student & 44 & 18.0\% \\ \cline{2-4} 
 & Private sector employee & 97 & 39.6\% \\ \cline{2-4} 
 & Public sector employee & 25 & 10.2\% \\ \cline{2-4} 
 & Self-employed & 29 & 11.8\% \\ \cline{2-4} 
 & Unemployed & 12 & 4.9\% \\ \cline{2-4} 
 & Housekeeping & 15 & 6.1\% \\ \cline{2-4} 
 & Retired & 6 & 2.4\% \\ \cline{2-4} 
 & Other & 17 & 6.9\% \\ \cline{1-4} 
\multirow{4}{*}{Family income} & Under 20000 \euro & 154 & 62.9\% \\ \cline{2-4} 
 & 20000-60000 \euro & 80 & 32.7\% \\ \cline{2-4} 
 & 60000-100000 \euro & 8 & 3.3\% \\ \cline{2-4} 
 & Over 100000 \euro & 3 & 1.2\% \\ \cline{1-4} 
\multirow{5}{*}{Marital status} & Married & 77 & 31.4\% \\ \cline{2-4} 
 & Divorced & 11 & 4.5\% \\ \cline{2-4} 
 & Single & 112 & 45.7\% \\ \cline{2-4} 
 & Widowed & 7 & 2.9\% \\ \cline{2-4} 
 & Domestic partnership & 38 & 15.5\% \\ \cline{1-4} 
\end{tabular}%
}
\end{table}
\medskip
\noindent\textbf{STTs Usage.} The majority of the participants (68.6\%) use STTs, such as wearable devices, for monitoring physical activity and physiological signals. Out of those, more than 70\% own a popular wearable brand, such as an Apple (21.4\%), Huawei (23.8\%), Samsung (17.9\%), or Garmin (8.9\%) device. They utilize their device to monitor indicators such as distance (48.9\%), steps (48.2\%), exercise duration (41.6\%), heart rate (40.8\%), calories (39.6\%), average time-per-kilometer (24.9\%), average speed (14.7\%), maximum speed (12.2\%), altitude (7.3\%), and water consumption (4.1\%).

The rationale behind the usage, or the lack thereof, of wearable devices by the users and non-users of our sample, respectively, varies (See Table\ref{tab:reasons}). With regard to the reasons behind the usage of wearable devices, more than one out of three participants use their wearable to gain more control over exercise (38.1\%), while a smaller percentage use them because it was gifted to them (16.7\%), for encouragement purposes (15.5\%), simply out of curiosity (11.9\%), because of interest in technological advancements (8.3\%), for health issues monitoring (6.0\%), through an insurance reward program (1.8\%), because of doctor's recommendations (1.2\%), or due to Covid-19 lockdowns (0.6\%). Concerning the reasoning behind non-usage, almost three out of four users do not own a device because of trust in their own body (27.3\%), high cost (22.1\%) or cannot pinpoint a reason (24.7\%). Fewer users report not owning a device because of distrust (7.8\%), lack of effectiveness (6.5\%), past negative experiences (5.2\%), lack of technological literacy (5.2\%), or lack of knowledge with regards to wearable devices (1.3\%).
\begin{table}[htb!]
\caption{Participants' responses regarding the reasons behind non-use or use of wearable devices\label{tab:reasons}}
\resizebox{\textwidth}{!}{%
\begin{tabular}{llll}
\hline
\textbf{Reason for non-usage} & \textbf{Percentage} & \textbf{Reason for usage} & \textbf{Percentage} \\ \hline 
I trust my own body & 27.3\% & More control over exercise & 38.1\% \\ \hline
I don't know & 24.7\% & Given as a present & 16.7\% \\ \hline
High cost & 22.1\%  & Encouragement purposes & 15.5\% \\ \hline
I do not trust them & 7.8\% & Curiosity & 11.9\% \\ \hline 
I don't need it & 6.5\% & Technology fan & 8.3\% \\ \hline
I had a negative experience in the past & 5.2\%  & Health issues monitoring & 6.0\% \\ \hline
Lack of technological literacy & 5.2\% & Insurance reward program & 1.8\% \\ \hline
Lack of knowledge w.r.t. wearables & 1.3\% & Doctor's recommendation & 1.2\% \\ \hline
& & Lockdown & 0.6\% \\ \hline
\end{tabular}%
}
\end{table}

Concerning their trust in the accuracy of the measured quantities, 93.5\% of the respondents show complete or some trust in the measuring accuracy. In contrast, only 3\% show distrust, and 3.6\% have no opinion on the matter. Regarding data privacy, almost half the respondents (44.6\%) have at least some concerns regarding privacy issues. However, the majority (53.0\%) still express no concern, and a small percentage (2.4\%) has no opinion on the matter. On a similar note, more than half the respondents engage in data sharing with friends (23.5\%), family (18.7\%), third-party platforms, such as Strava (6.6\%), family doctor (3.6\%), or social media in general (1.8\%).

\begin{table}[htb!]
\caption{Mean ($\mu$) and Standard Deviation ($\sigma$) values, and Cronbach's alpha ($\alpha$) for each dimension of the negotiation strategies in StoryWear. All dimensions show near-acceptable to excellent internal consistency. Results on the item-correlation test are also presented and show medium to high coherence between an item and the other items in a test.\label{tab:validity}}
\resizebox{\textwidth}{!}{%
\begin{tabular}{llllll}
\hline
\textbf{Negotiation Dimensions} & \textbf{$\mu$} & \textbf{$\sigma$} & \textbf{$\alpha$} & \textbf{\makecell{Range of\\ $\alpha$ if items\\ deleted}} & \textbf{\makecell{Range of\\ corrected item\\ total correlations}}\\ \hline
Knowledge & 3.93 & 1.50 & .804 & .967 & .499-.574\\ \hline
Negative Impact Understanding by Non-Exercise & 4.44 & 1.38 & .683 & .966-.968 & .431-.660\\ \hline
Self-motivation & 4.87 & 1.60 & .832 & .966-.967 & .589-.736\\ \hline
Enable & 4.04 & 1.72 & .877 & .966 & .679-.767\\ \hline
Socialization & 3.78 & 1.85 & .904 & .966 & .696-.718\\ \hline
Enhancement & 4.33 & 1.62 & .787 & .966-.967 & .620-.726\\ \hline
Commitment & 4.63 & 1.61 & .865 & .966 & .711-.755\\ \hline
Create Pulse & 3,93 & 1.62 & .804 & .966 & .633-.674\\ \hline
Encouragement Relations Development & 3.70 & 1.70 & .832 & .966 & .685-.706\\ \hline
Time & 4.18 & 1.77 & .907 & .966 & .709-.743\\ \hline
Financial & 3.80 & 1.86 & .913 & .966 & .689-.752\\ \hline
\end{tabular}%
}
\end{table}
\medskip
\noindent\textbf{StoryWear Internal Consistency Reliability Assessment.}
To ensure the stability and truthfulness of our findings, we report here common indicators for model reliability assessment. Specifically, we used Cronbach's alpha ($\alpha$) \cite{cronbach1951coefficient} to test the reliability of the scale for the 33 items, revealing excellent internal consistency ($\alpha=0.967$). Additionally, we utilized the Corrected Item-Total Correlation test to evaluate the coherence between an item and the other items in the scale and ultimately verify the reliability of StoryWear. Table~\ref{tab:validity} (column Range of corrected item-total correlations) shows the correlation (Pearson's r) between each item and a scale score that excludes that item. The r-values range between .431 and .767, indicating once again high internal consistency ($r>.3$) \cite{de2002analyzing}. Also, in the same table (column Range of $\alpha$ if items deleted), we notice that there are no items that, if deleted, would lead to a ``substantial'' increase in Cronbach's $\alpha$. If there were, then the specific items could be discarded; hence we highlight the necessity of all included items. With regards to the internal consistency of the variables (assuming the original eleven negotiation dimensions), the alpha indicators showed acceptable to excellent values, ranging over 0.70 \cite{churchill1979paradigm,hinkin1998brief} for ten out of eleven dimensions. One dimension (negative impact understanding by non-exercise) showed a slightly lower score, which, although weaker, suffices for early stages of research \cite{nunnally1994psychometric}. Table~\ref{tab:validity} presents the indicators and values resulting from the internal consistency analysis. 

\medskip
\noindent\textbf{\textit{Overall effectiveness of Negotiation Strategies.}} Our results, as presented in Table~\ref{tab:validity}, indicate differences in the negotiation dimensions' perceived effectiveness by our sample. Specifically, self-motivation ($\mu=4.87$ and $\sigma=1.60$), commitment ($\mu=4.63$ and $\sigma=1.61$), negative impact of non-exercise understanding ($\mu=4.44$ and $\sigma=1.38$), enhancement ($\mu=4.33$ and $\sigma=1.62$), time ($\mu=4.18$ and $\sigma=1.77$), and enable ($\mu=4.04$ and $\sigma=1.72$) -listed in decreasing order of perceived effectiveness- emerged as the most preferred dimensions ($\mu>4.0$). While knowledge ($\mu=3.93$ and $\sigma=1.50$), create pulse ($\mu=3.93$ and $\sigma=1.62$), financial ($\mu=3.80$ and $\sigma=1.86$), socialization ($\mu=3.78$ and $\sigma=1.85$), and encouragement relations development ($\mu=3.70$ and $\sigma=1.70$), have near neutral score (i.e., 3.5).

\begin{table}[htb!]
\caption{The results of the independent samples t-test and cross-tabulation (indicated as a t-score), and the one-way ANOVA (indicated as an F-score). Statistically significant results ($p<0.05$) are bold and colored in purple for readability. Each column corresponds to a user characteristic (independent factors), and each row corresponds to a negotiation dimension (dependent factor).\label{tab:factors}}
\resizebox{\textwidth}{!}{%
\begin{tabular}{lllllll}
\hline
\multicolumn{1}{c}{\textit{\textbf{Dependent Factor}}} & \multicolumn{6}{c}{\textit{\textbf{Independent Factors}}} \\ \hline
\multicolumn{1}{c}{\textbf{Negotiation Strategies Dimension}} & \multicolumn{1}{c}{\textbf{Gender}} & \multicolumn{1}{c}{\textbf{Age}} & \multicolumn{1}{c}{\textbf{Marital Status}} & \multicolumn{1}{c}{\textbf{Education Level}} & \multicolumn{1}{c}{\textbf{Occupation}} & \multicolumn{1}{c}{\textbf{Wearables Usage}} \\ \hline
Knowledge & t = 0.622 & t = 0.017 & F = 1.031 & F = 1.558 & F = 0.453 & {\color[HTML]{4B0082} \textbf{t = 5.224}} \\ \hline
Negative Impact of Non-Exercise Understanding & t = 0.392 & t = 0.474 & F = 0.836 & F = 1.771 & F = 1.521 & {\color[HTML]{4B0082} \textbf{t = 3.419}} \\ \hline
Self-motivation & {\color[HTML]{4B0082} \textbf{t = 2.110}} & {\color[HTML]{4B0082} \textbf{t = 2.303}} & F = 2.237 & F = 0.905 & F = 1.286 & {\color[HTML]{4B0082} \textbf{t = 2.459}} \\ \hline
Enable & t = 0.793 & t = 0.400 & F = 0.509 & F = 0.670 & F = 0.864 & {\color[HTML]{4B0082} \textbf{t = 2.257}} \\ \hline
Socialization & t = 0.772 & t = 1.412 & F = 1.092 & F = 0.163 & F = 0.425 & t = 1.345 \\ \hline
Enhancement & t = 0.015 & t = 0.685 & F  = 1.398 & F = 1.430 & F = 0.497 & {\color[HTML]{4B0082} \textbf{t = 3.782}} \\ \hline
Commitment & t = 1.156 & t = 0.409 & F  = 1.142 & F = 0.581 & F = 0.558 & {\color[HTML]{4B0082} \textbf{t = 2.992}} \\ \hline
Create Pulse & t = 0.391 & t = 1.602 & F = 0.422 & F = 1.281 & F = 0.307 & {\color[HTML]{4B0082} \textbf{t = 3.593}} \\ \hline
Encouragement Relations Development & t = 1.353 & t = 0.684 & F = 0.897 & F = 0.301 & F = 0.240 & {\color[HTML]{4B0082} \textbf{t = 2.550}} \\ \hline
Time & t = 0.120 & t = 1.171 & F = 1.531 & {\color[HTML]{4B0082} \textbf{F = 2.725}} & F = 0.628 & {\color[HTML]{4B0082} \textbf{t = 2.522}} \\ \hline
Financial & t = 0.152 & {\color[HTML]{4B0082} \textbf{t = 2.293}} & F = 0.192 & F = 1.860 & F = 0.420 & t = 1.523 \\ \hline
\end{tabular}%
}
\end{table}
\noindent\textbf{\textit{Effect of User-related Factors on Negotiation Strategies.}} To explore the effects of various user-related factors, such as gender, age, occupation, and STTs' usage, among others, we perform independent samples t-test and cross-tabulation and the one-way ANOVA analysis. The results (i.e., t-score for t-test and F-score for ANOVA), as shown in Table~\ref{tab:factors}, indicate the independent factors of marital status and occupation do not affect the negotiation strategies' perceived effectiveness. Similarly, gender has a significant effect ($p<0.05$) only on the self-motivation dimension of negotiation strategies ($\mu_{male}=5.11$ and $\mu_{female}=4.68$), while education level only has a significant effect ($p<0.05$) on the time dimension ($\mu_{vocational}=3.47$ and $\mu_{university}=4.36$). Age has a statistically significant effect ($p<0.05$) on the self-motivation ($\mu_{17-39}=5.03$ and $\mu_{40-59}=4.54$) and financial ($\mu_{17-39}=3.99$ and $\mu_{40-59}=3.42$) dimensions. Contrary to other user-related factors, the usage of STTs, specifically wearable devices, has a statistically significant effect ($p<0.05$) on the majority of the negotiation dimensions, excluding the socialization and financial dimensions, indicating the importance of this parameter. 

\begin{figure}[htb!]
  \centering
  \includegraphics[width=\textwidth]{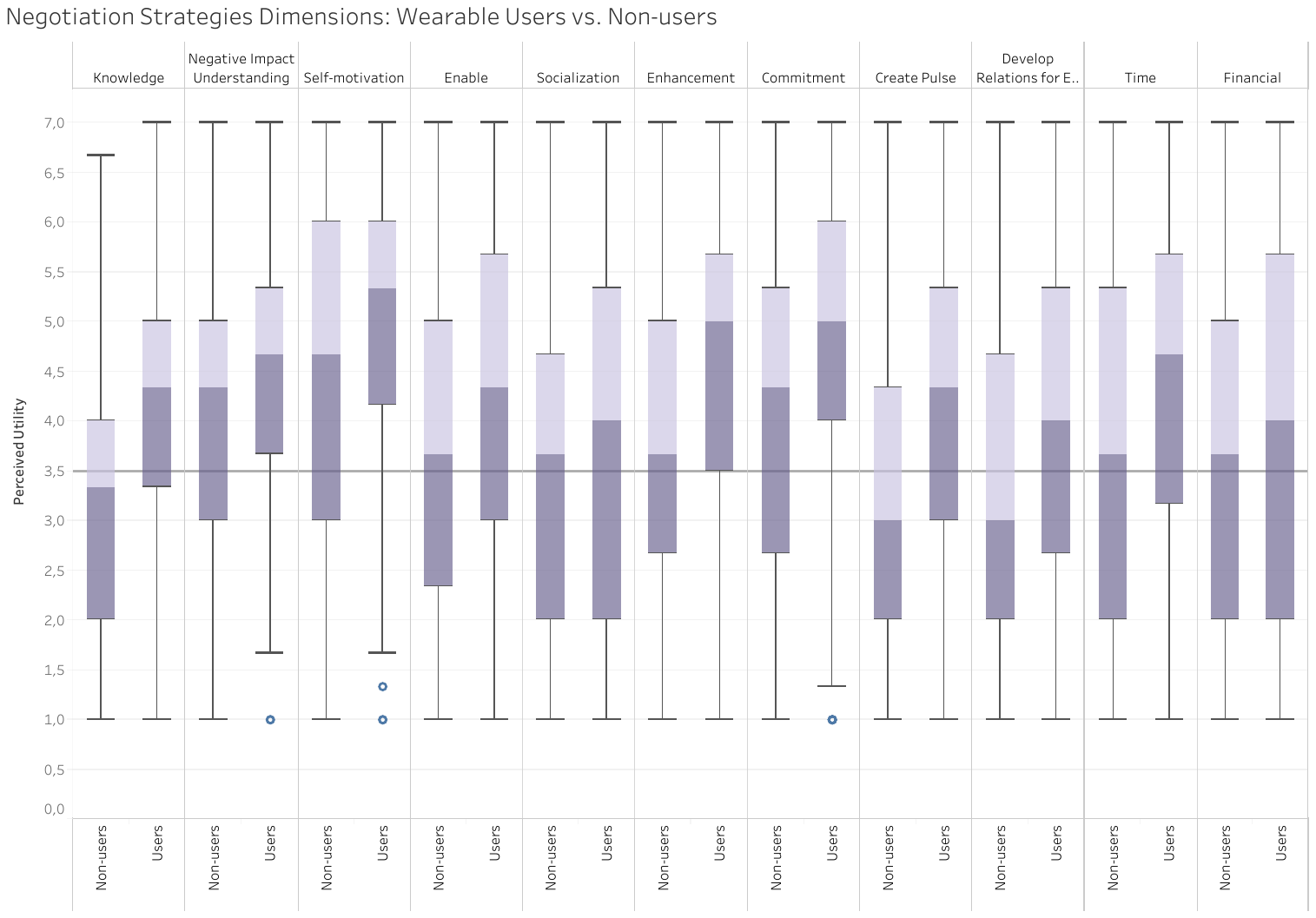}
  \caption{Boxplot comparison of negotiation strategies' perceived effectiveness by STTs' users and non-users.\label{fig:dimensions}}
\end{figure}
\noindent\textbf{\textit{Differences in effectiveness of Negotiation Strategies between Users and Non-users.}}
As discussed above, STT users see higher effectiveness in negotiation features compared to non-users. However, they both rank the top negotiation dimensions similarly, with self-motivation features having the highest mean for both groups ($\mu_{users}=5.05$ and $\mu_{non-users}=4.46$), followed by commitment ($\mu_{users}=4.84$ and $\mu_{non-users}=4.16$), and negative impact of non-exercise understanding  ($\mu_{users}=4.64$ and $\mu_{non-users}=4.01$) features. 
However, some high-performing negotiation strategies ($\mu>4$) for users, such as ``Knowledge'' ($\mu_{users}=4.26$) and ''Create Pulse'' ($\mu_{users}=4.17$), are rated below average by non-users ($\mu_{non-users}=3.23$ and $\mu_{non-users}=3.39$, respectively) with regards to perceived effectiveness, with a difference of approximately one unit. Figure~\ref{fig:dimensions} shows the distribution and skewness of users and non-users' perceived effectiveness of different negotiation dimensions.

\section{Discussion\label{discussion}}
This section discusses our findings and explores how the inclusive StoryWear scale can be applied in practice for incorporating negotiation strategies into the design of SSTs to encourage fitness and well-being. To this end, it provides design implications and actionable guidelines with regard to negotiation strategies appealing to a broader audience or tailored to specific users.

Our internal consistency reliability assessment (See Table~\ref{tab:validity}) demonstrates the high reliability of our scale, StoryWear, and enables its usage in future research. It also indicates that \textbf{StoryWear manages to translate the abstract concepts of the Transtheoretical Model and the Leisure Constraints Negotiation Model into concrete STTs' features} (i.e., negotiation features), captured through its inclusive storyboards format.

With regards to the overall perceived utility of negotiation strategies translated into STTs' features, we notice that participants generally recognize the effectiveness of negotiation features in overcoming constraints regarding exercise (See Table~\ref{tab:validity}). Specifically, they prioritize features that encourage them to reassess their preconceptions regarding their physical activity and well-being or understand the rationale behind exercise participation (e.g., health benefits, physical appearance improvement, confidence boost, sedentarism decrease). Additionally, they value features that enable them to understand the negative impact of non-exercise on their body and mind and features that support them in committing to exercise through realistic goals and training plans.

Breaking down these results by user segment (based on demographics and STT usage), we highlight commonalities and differences across groups and negotiation features. The differences in the self-motivation dimension showcase that men (as compared to women), and participants in initial adulthood (17-39) (as compared to middle and final adulthood), are more interested in the relationship between exercise and health, physical appearance, and confidence, and the respective negotiation features. With regards to the financial dimension, we see that participants in initial adulthood see higher effectiveness in budgeting features, which may be because adults in middle or final adulthood are more financially stable\footnote{``US household wealth rose to record \$141.7 trillion in Q2, Fed says.'' 23 Sept. 2021, \url{https://www.reuters.com/business/finance/us-household-wealth-rose-record-1417-trillion-q2-fed-says-2021-09-23/}. Accessed 8 Nov. 2021.}. Finally, when it comes to the time dimension, our findings indicate that university graduates are more appreciative of features that help them organize their daily schedule to account for exercise compared to vocational school graduates, which may be due to differences in routine or interest in exercise itself. 

However, STT usage was the most separative factor, as \textbf{there is a significant difference between users and non-users of STTs in the way they perceive the effectiveness of the majority of the negotiation strategies} translated into technological features, except for the socialization and financial dimensions.
We conjecture that this is because the scale distribution coincided with the Covid-19 lockdown and restriction measurements, which may have impacted the latter dimensions. Specifically, all paid forms of exercise, such as gyms and classes, were suspended until further notice. Only free forms of exercise, such as walking or running outdoors, were available to participants, which may have lowered the overall effectiveness of financial negotiation features. Similarly, group training or even exercise buddies were prohibited due to social distancing, which we assume might have impacted the perceived effectiveness of the socialization features. 

Overall, our findings demonstrate the effectiveness of negotiation strategies for STTs' design regarding encouraging physical activity and well-being. We have proven that certain strategies and their respective negotiation features have higher perceived effectiveness than others. In the following, we provide a series of concrete guidelines on how each strategy could be translated into relevant negotiation features beyond our representation. 

\textit{Self-motivation} is the most useful of all negotiation strategies, as per both wearables' users and non-users. Therefore, to appeal to a broader user base, STTs promoting fitness and well-being should be designed to allow users to challenge their preconceptions regarding exercise. This can be achieved through \textbf{system features that showcase cause-and-effect linkage and projected outcome of a user's health behavior}, such as real-time or analytical feedback, AI-based coaching, personalized training plans, and content, or cause-and-effect metaphors (i.e., growing gardens, virtual pets that bloom/grow with physical activity and wither with inactivity). Such features enable users to understand how a positive behavior change can enhance their health, confidence, and appearance to reassess and reaffirm their relationship with exercise. 

Similarly, \textit{commitment} shows high perceived effectiveness regardless of the usage of STTs. Hence, it has the potential to appeal to the broader user base. Hence, ubiquitous systems for self-tracking should incorporate \textbf{features that reinforce the user's sense of commitment to physical activity}. Such features include personalized and adaptive goal-setting functionality, multiple simultaneous goals, or graded goals and difficulty levels, personal and group challenges and one-to-one competitions, exercise buddy functionality, or social sharing features for taking advantage of the user's need to maintain a certain public image regarding exercise with their network. For a user to commit to a health behavior, the behavior has to be challenging enough to motivate the user but not unrealistically demanding to be deemed discouragingly impossible.

Our results also show that \textit{understanding of the negative impact of non-exercise} elicited positive effectiveness feedback from our entire user sample, rendering it a good candidate for implementation in a real-world system for the general public. To exercise, individuals employ negotiation strategies related to awareness of the negative consequences of non-participation, such as the health problems that can arise, to reassess personal values for participation to improve their physical appearance and overall health, and to prevent the negative health effects of sedentarism and inactivity. They also use exercise to deal with everyday inactivity and hypomobility, for example, to avoid staying inactive at home, and they make sure to include participation in exercise programs in their weekly schedule. Positive cognitive and affective attitudes refer to perceptions of the beneficial effects of exercise and the emotions that develop through participation. Urban green spaces for recreation are potential factors for creating positive affective attitudes towards physical activity. Participation in outdoor recreation activities is associated with positive experiences \cite{stevens2019social} and evoking emotions that promote personal well-being \cite{filo2016exploring}. Modern lifestyles and alienation from the natural environment make it imperative to promote outdoor exercise programs. Ubiquitous STTs promoting physical activity and well-being should be \textbf{designed to raise awareness regarding the health hazards of physical inactivity} through cause-and-effect linkage for negative behaviors, negative feedback, or reward loss for failing to achieve set goals and information provision through dedicated, trustworthy channels. However, features that have a punitive character should be used with caution on a case-by-case basis, as for some users, praise might lead to better results than scolding.

On a smaller scale, \textit{enhancement}, \textit{time}, and \textit{enable} also elicited positive responses from the participants overall. Hence, STTs for promoting physical activity should incorporate \textbf{features that help users incorporate exercise into their daily routine, as a means of tension relief or sedentariness and inactivity prevention}, as well as \textbf{features for rewarding them for doing so}. These negotiation dimensions could take the form of just-in-time stress-relief mechanisms, physical and mental self-care functionality, intrinsic and extrinsic reward schemes, and contextualized exercise goals and plans based on the user's schedule and daily routine.

Unlike the strategies above that are considered helpful by STTs' users and non-users alike, the \textit{create pulse} and \textit{knowledge} strategies seem to have higher appeal to STTs' users compared to non-users. Hence, negotiation features based on these strategies may be less preferred for attracting new users. However, for the existing, engaged user base, such \textbf{features should be designed to create external stimuli for encouraging physical activity beyond existing internal motivation}, as well as \textbf{enabling the experienced user to gain additional knowledge regarding exercise}. These negotiation dimensions could take the form of information provision, tailored content promotion, and just-in-time reminders related to exercise, among others.

On the contrary, both users and non-users seem to find the \textit{socialization}, \textit{financial} and \textit{encouragement relations development} dimensions less useful. As mentioned above, this low perceived effectiveness might be time-dependent due to the Covid-19 restrictions, especially in terms of social distancing and fitness and wellness businesses' suspension of operation. Alternatively, people who have decided to exercise have probably already taken care of the financial burden or are not interested in people's company or social acceptance and support from their environment because they believe their decision is beneficial for their health. So they use these strategies to a very small extent. It is interesting to note here that while social features seem to have lower appeal to participants, they are available in most commercial self-tracking devices, contrary to some negotiation features with high perceived effectiveness, such as self-motivation and understanding of the negative impact of non-exercise, which are rarely encountered in such devices.

Participation in recreational activities ``does not depend on the absence of constraints (although this may be true for some people) but on negotiating through them, and such negotiations can modify rather than exclude participation'' \cite{jackson1993negotiation}. This proposition entails that all individuals perceive some type of constraint and that the final decision (to commit or not to commit) is not always inclusive. In addition, constraints may lead to a modified commitment (e.g., less time or choosing an alternative activity).
If we take into account the changes in people's desires and preferences for the various activities and the psychophysiological part of recreation and combine them with their leisure time and their style of exercise selection, then we can define the guidelines concerning the maintenance, planning, utilization and development of the various recreational constraints.

\section{StoryWear Implications\label{uptake}}
\noindent\textbf{Theoretical Implications.} Our results highlight that the usage of STTs positively affects the perceived effectiveness of negotiation strategies for overcoming leisure constraints. A technologist's strength is to see the big picture and understand why and when some techniques work and others fail. To this end, this research highlights the benefits of theoretically-founded design and points to a future of specialized, personalized training services tailored to the individual's needs and aiming at maximizing training performance and goal accomplishment. For instance, research supports that the difficulty level of goals and challenges set by the system for the user needs to correspond to the abilities of the respective user. This way, user engagement is maximized because trivial goals can cause boredom, while over-demanding goals can cause tension and stress. Such negative feelings are not in line with the concept of recreation and boost the effect of constraints concerning exercise participation. Hence, as mentioned in Section \ref{introduction}, it is important to properly utilize psychological and behavioral theories in designing and implementing fitness self-tracking apps to encourage user engagement and repeat participation in physical activity.

Every user has the right to personalized training, which, in the era of self-tracking devices, is more feasible and affordable than ever. Such a paradigm shift significantly changes the role of personal trainers and coaches, who can now monitor their clients' data originating from STTs, evaluate their performance, and recommend necessary -and informed- changes to their exercise routine for the users to achieve their training goals. STTs nowadays go beyond self-tracking to motivation and accountability mechanisms. Sports scientists and HCI designers ought to stay up to date with the advancements in the domain and listen to the users' needs. However, technology is no panacea, and a hybrid training model is still preferable. For instance, a gamification intervention introduced in a physical fitness class can motivate the users, increasing user interest, goal accomplishment, and commitment. At the same time, a physical trainer can ensure the proper performance of the exercises to avoid potential injuries. 
\bigbreak
\noindent\textbf{Practical Implications.} Having explained the theoretical importance of our work, we provide two indicative uptake scenarios to showcase the practical effectiveness of the StoryWear scale for ubiquitous computing and HCI researchers and practitioners. 

For instance, to promote scientific advancement, a group of Informatics researchers with scientific interests surrounding ubiquitous computing and HCI is interested in conducting an intervention to encourage adolescents to exercise more through the use of wearable technology. To this end, they can utilize the current article and the StoryWear scale to broaden their interdisciplinary knowledge in the intersection between Sports Science, Behavioral Science, and Computer Science. Following, they can implement a field study utilizing StoryWear to collect user feedback to explore how ubiquitous HCI design incorporating negotiation strategies can be tailored to their targeted user group, i.e., adolescents. After all, the novel, creative format of StoryWear has better chances of appealing to their young audience compared to traditional means. Hence, having completed their study, they can now design an intervention with theoretically sound, user-centric components based on their target users' most effective negotiation strategies. 

Concerning real-life adoption, a city's committee on aging is interested in taking action against the increased sedentariness of the city's elderly population utilizing recent technological advancements in ubiquitous computing. However, given the low technological literacy of the city's elderly population, they would like to evaluate which negotiation features might be more successful in encouraging their target users to perform physical activity. To this end, they can utilize StoryWear in combination with user interviews to elicit the users' feedback on the perceived effectiveness of the various wearable components. StoryWear's easy-to-understand, visual format offers a more straightforward understanding than textual descriptions incorporating technical terms. Thus, they can now decide upon the most appropriate negotiation features to implement, cutting down on unnecessary implementation costs and maximizing the effectiveness of their intervention.

\section{Conclusions, Limitations \& Future Work} \label{conclusions}
To the best of the authors' knowledge, the current work explores the relationship between the Leisure Constraints Negotiation Model and ubiquitous STTs for the first time in literature (C1). Specifically, the paper contributes to the HCI and wearables community by advancing the understanding of how different negotiation strategies can be translated into and implemented as ubiquitous negotiation features and how the usage of STTs affects the negotiation strategies' perceived effectiveness. In this work, we have introduced and validated a new scale, StoryWear, representing the negotiation features above in an inclusive, easy-to-understand storyboard format (C2). Through achieving satisfactory internal consistency for all StoryWear's dimensions, we have verified that the negotiation features and their novel presentation format successfully represent the original negotiation strategies scale. To this end, we provide concrete guidelines on applying the Leisure Constraints Negotiation Model in practice for STTs' design (C3).

Our findings indicate that self-motivation, commitment, and the negative impact of non-exercise understanding are the most successful negotiation strategies overall, regardless of the self-tracking practices of the respondents. However, users of STTs for exercise and well-being express significantly higher perceived effectiveness for the negotiation features, excluding dimensions that may depend on external factors, such as socialization and financial dimensions, where both parties showed near-neutral perceived effectiveness. Additionally, non-users show below-average perceived effectiveness for features belonging to the knowledge, create pulse, and socialization dimensions. Our results could guide practitioners, researchers, and ubiquitous designers in incorporating and prioritizing negotiation strategies into the design of self-tracking applications to enable users to negotiate through the various obstacles of behavior adoption. \rev{It is important to note here that while design guidelines can facilitate the development of STTs, researchers and practitioners need to be mindful of their epistemic status. Specifically, while we strongly believe in the value of empirical data for generating design guidelines, we recommend that our guidelines be treated similarly to “design hypotheses”, which require additional testing rather than hard truths.}

While our discussion elaborates on the contributions of our work and its outputs, there are certain limitations that we need to address. First, our study presents the perceived effectiveness of various negotiation strategies translated into ubiquitous negotiation features and implemented in storyboards. However, the actual effectiveness of the strategies may differ when implemented in a real-world intervention. Therefore, we plan to experiment with a number of negotiation features to evaluate their real-world effectiveness captured through changes in the users' physical activity and sedentariness levels. Moreover, while we have validated the StoryWear scale in the physical activity and well-being domain and can claim the concept's applicability in other health behavior domains, such as smoking cessation, or decrease in alcohol consumption, our findings and scale should be utilized after appropriate adaptation catering to the particularity of each domain. Similarly, even though we have validated StoryWear through three distinct, small-scale preliminary studies and user samples, the utilization of the scale for different cohorts should be undertaken with caution, having ensured the understandability and reliability of the scale for each specific cohort. For example, during the questionnaires' distribution, we noticed that adults in the final adulthood stage (60+), when technologically illiterate, had trouble understanding the concepts described in the storyboards. To this end, we intend to organize participatory design workshops to elicit feedback from more demanding cohorts and implement a large-scale study recruiting diverse user cohorts. Also, we intend to repeat our study in the after-Covid-19 era to evaluate the potential differences in the perceived effectiveness of negotiation features. Finally, had we chosen to represent the negotiation strategies differently (i.e., translating them into features other than the ones presented in StoryWear), we might have elicited different responses of self-reported effectiveness, corresponding to the same high-level negotiation dimension. To cover this gap, we provide alternative negotiation feature ideas corresponding to all negotiation strategies in the discussion above.

Summing up, this work presents many insightful and significant findings and opens up many areas for further exploration. Future work should focus on large-scale validation of the proposed scale, StoryWear, as well as validation in minority user samples or populations in different cultural settings or speaking different languages. Additionally, while this work has explored the effect of demographics and self-tracking adoption on the perceived effectiveness of negotiation features, future research can compare the effectiveness of such features depending on alternative factors, such as personality type \cite{goldberg1990alternative}, stage of behavior change \cite{prochaska1997transtheoretical}, or stage of lived informatics \cite{rooksby2014personal} of the respondent. Finally, our results should be validated in other health behavior domains, such as smoking cessation, discouraging drug use, or diet monitoring, to investigate possible changes in the perceived effectiveness of the negotiation features.

\section*{Declarations}
This project has received funding from the European Union’s Horizon 2020 research and innovation programme under the Marie Skłodowska-Curie grant agreement No 813162. The content of this paper reflects only the authors' view and the Agency and the Commission are not responsible for any use that may be made of the information it contains. The authors would like to thank A. Lykoglou for her creativity and artistic contribution in creating the storyboards, and DUTH students and graduates, P. Masiou, D. Tsilikas, and D. Georgopoulou, for their invaluable field presence and the distribution of the StoryWear scale. The authors declare no competing interests. 

\section*{Data Availability}
The datasets generated during and/or analysed during the current study are available from the corresponding author on reasonable request.

\begin{appendices}

\section{From Negotiation Strategies to Negotiation Features\label{ap:features}}
This appendix introduces the mapping between negotiation strategies and our negotiation features which are the foundation of our storyboards scale, StoryWear (See Table~\ref{tab:mapping}).

\setcounter{table}{5}
\renewcommand{\thetable}{\arabic{table}}
\begin{longtable}{|p{.5\linewidth}|p{.5\linewidth}|}
\caption{The 33 negotiation strategies translated into equivalent negotiation features.\label{tab:mapping}}\\ \hline
\multicolumn{1}{|c|}{\textbf{Negotiation Strategies}} & \multicolumn{1}{c|}{\textbf{Negotiation Features}} \\ \hline
\multicolumn{2}{|c|}{\textit{1. Knowledge}} \\ \hline
I read books about exercise in an attempt to learn more about it & Notifications about new book publications w.r.t. exercise \\ \hline
I read articles about exercise in an attempt to learn more about it & Notifications about new online articles w.r.t. exercise \\ \hline
I engage in discussions in an attempt to learn more about exercise & Subscription to fitness influencers' content updates \\ \hline
\multicolumn{2}{|c|}{\textit{2. Negative Impact of Non-Exercise Understanding}} \\ \hline
Warnings about health hazards of inactivity move me emotionally & Warnings about health hazards of inactivity coupled with exercise recommendations \\ \hline
Dramatic portrayals of the evils of inactivity move me emotionally & Personalized, dramatic warnings about health hazards based on my fitness history \\ \hline
I react emotionally to information about health hazards of inactivity & Chat-based information provision about the health hazards of physical inactivity \\ \hline
\multicolumn{2}{|c|}{\textit{3. Self-motivation}} \\ \hline
I am considering the idea that regular exercise would make me a healthier person & Timeline visualization of fitness indicators improvement showcasing progress \\ \hline
I consider the fact that I would feel more confident in myself if I exercised regularly & Real-time monitoring and reinforcement feedback on exercise \\ \hline
I consider the fact that my appearance would be better if I exercised regularly & Awareness visualization of exercise effects on the user's body and appearance \\ \hline
\multicolumn{2}{|c|}{\textit{4. Enable}} \\ \hline
Instead of remaining inactive, I engage in some physical activity & Just-in-time movement reminders after prolonged sedentariness \\ \hline
Rather than viewing exercise as simply another task to get out of the way, I try to use it as my special time to relax and recover from the day's worries & Enjoyable exercise recommendations for self-care \\ \hline
When I am feeling tense, I find exercise a great way to relieve my worries & Stress detection feature accompanied with relaxation recommendations \\ \hline
\multicolumn{2}{|c|}{\textit{5. Socialization}} \\ \hline
I try to find people to exercise with & Group training planning for finding exercise buddies \\ \hline
I try to hang out with people who exercise & Collective, group-based exercise goals functionality \\ \hline
I try to persuade my friends to exercise together & One-to-one exercise invitations \\ \hline
\multicolumn{2}{|c|}{\textit{6. Enhancement}} \\ \hline
I do something nice for myself for making efforts to exercise more & ``Guilty pleasure'' reward for achieving exercise goals \\ \hline
I try to set realistic goals for myself rather than setting myself up for failure by expecting too much & Multi-modal goal setting with gradual goal accomplishment possibility \\ \hline
I reward myself when I exercise & Graded virtual rewards for exercise goal accomplishment \\ \hline
\multicolumn{2}{|c|}{\textit{7. Commitment}} \\ \hline
I tell myself I am able to keep exercising if I want to & Realistic goal-setting and exercise recommendations based on user fitness profile \\ \hline
I tell myself that if I try hard enough, I can keep exercising & Personal challenges, and goal-setting functionality \\ \hline
I make commitments to exercise & Adaptive goal-setting and public commitments through social sharing functionality \\ \hline
\multicolumn{2}{|c|}{\textit{8. Create Pulse}} \\ \hline
I put things around my home to remind me of exercising & Just-in-time notifications for equipment pick up \\ \hline
I buy sports equipment & New fitness products notifications from popular stores \\ \hline
I buy sports clothes and accessories & Easy clothes sizing guide tool based on measured indicators \\ \hline
\multicolumn{2}{|c|}{\textit{9. Encouragement Relations Development}} \\ \hline
I try to have a healthy friend who encourages me to exercise when I do not feel up to it & Progress monitoring and feedback for close friends functionality \\ \hline
I try to find someone who provides feedback about my exercising & Virtual or real digital trainer functionality \\ \hline
I try to find someone with similar interests who provides support about my exercise & Social media connectivity for finding people with similar interests for seeking support \\ \hline
\multicolumn{2}{|c|}{\textit{10. Time}} \\ \hline
I try to manage my time more wisely so that I have time to exercise & Just-in-time notifications to exercise to help users stick to their plan \\ \hline
I plan out my work and responsibilities so that I have time to exercise & Contextualized notifications (based on work and routine) to encourage users to exercise \\ \hline
I use part of my free time to exercise & Personalized exercise recommendations of varied lengths and activity types \\ \hline
\multicolumn{2}{|c|}{\textit{11. Financial}} \\ \hline
I try to budget my money to cover the cost of exercise & Recommendations of healthy-vs-unhealthy purchases' swaps for smarter budgeting \\ \hline
I have just learned to live within my means to cover the cost of exercise & Virtual wallet for saving up for exercise \\ \hline
I engage in financial planning to manage the cost of exercise & Contextualized visualization of exercise availability and prices in the user's vicinity for planning ahead \\ \hline
\end{longtable}%
\end{appendices}

\bibliography{sn-bibliography}%

\begin{thebibliography}{85}
\providecommand{\natexlab}[1]{#1}
\providecommand{\url}[1]{{#1}}
\providecommand{\urlprefix}{URL }
\providecommand{\doi}[1]{\url{https://doi.org/#1}}
\providecommand{\eprint}[2][]{\url{#2}}
 \bibcommenthead

\bibitem[{Adams et~al(2015)Adams, Costa, Jung, and
  Choudhury}]{adams2015mindless}
Adams AT, Costa J, Jung MF, et~al (2015) Mindless computing: designing
  technologies to subtly influence behavior. In: Proceedings of the 2015 ACM
  international joint conference on pervasive and ubiquitous computing, pp
  719--730

\bibitem[{Ajzen(1991)}]{ajzen1991theory}
Ajzen I (1991) The theory of planned behavior. Organizational behavior and
  human decision processes 50(2):179--211

\bibitem[{Aldenaini et~al(2020)Aldenaini, Alqahtani, Orji, and
  Sampalli}]{aldenaini2020trends}
Aldenaini N, Alqahtani F, Orji R, et~al (2020) Trends in persuasive
  technologies for physical activity and sedentary behavior: a systematic
  review. Frontiers in artificial intelligence 3:7

\bibitem[{Alexandris et~al(2002)Alexandris, Tsorbatzoudis, and
  Grouios}]{alexandris2002perceived}
Alexandris K, Tsorbatzoudis C, Grouios G (2002) Perceived constraints on
  recreational sport participation: Investigating their relationship with
  intrinsic motivation, extrinsic motivation and amotivation. Journal of
  Leisure research 34(3):233--252

\bibitem[{Alexandris et~al(2007)Alexandris, Kouthouris, and
  Girgolas}]{alexandris2007investigating}
Alexandris K, Kouthouris C, Girgolas G (2007) Investigating the relationships
  among motivation, negotiation, and alpine skiing participation. Journal of
  Leisure Research 39(4):648--667

\bibitem[{Balaska et~al(2019)Balaska, Yfantidou, Kenanidis, Spyridopoulou, and
  Alexandris}]{balaska2019}
Balaska P, Yfantidou G, Kenanidis T, et~al (2019) Exploring how recreational
  sport participants with different motivation levels use leisure negotiation
  strategies. In: European Academy of Management, EURAM 2019

\bibitem[{Bandura(1999)}]{bandura1999social}
Bandura A (1999) Social cognitive theory of personality. Handbook of
  personality 2:154--96

\bibitem[{Bentvelzen et~al(2021)Bentvelzen, Niess, Wo{\'z}niak, and
  Wo{\'z}niak}]{bentvelzen2021development}
Bentvelzen M, Niess J, Wo{\'z}niak MP, et~al (2021) The development and
  validation of the technology-supported reflection inventory. In: Proceedings
  of the 2021 CHI Conference on Human Factors in Computing Systems, pp 1--8

\bibitem[{Busch et~al(2016)Busch, Mattheiss, Reisinger, Orji, Fr{\"o}hlich, and
  Tscheligi}]{busch2016more}
Busch M, Mattheiss E, Reisinger M, et~al (2016) More than sex: The role of
  femininity and masculinity in the design of personalized persuasive games.
  In: International Conference on Persuasive Technology, Springer, pp 219--229

\bibitem[{Cacioppo and Petty(1984)}]{cacioppo1984elaboration}
Cacioppo JT, Petty RE (1984) The elaboration likelihood model of persuasion.
  ACR North American Advances

\bibitem[{Caraban et~al(2020)Caraban, Konstantinou, and
  Karapanos}]{caraban2020nudge}
Caraban A, Konstantinou L, Karapanos E (2020) The nudge deck: A design support
  tool for technology-mediated nudging. In: Proceedings of the 2020 ACM
  Designing Interactive Systems Conference, pp 395--406

\bibitem[{Chen et~al(2019)Chen, Li, Rosner, and
  Hiniker}]{chen2019understanding}
Chen YY, Li Z, Rosner D, et~al (2019) Understanding parents' perspectives on
  mealtime technology. Proceedings of the ACM on Interactive, Mobile, Wearable
  and Ubiquitous Technologies 3(1):1--19

\bibitem[{Churchill~Jr(1979)}]{churchill1979paradigm}
Churchill~Jr GA (1979) A paradigm for developing better measures of marketing
  constructs. Journal of marketing research 16(1):64--73

\bibitem[{Clawson et~al(2015)Clawson, Pater, Miller, Mynatt, and
  Mamykina}]{clawson2015no}
Clawson J, Pater JA, Miller AD, et~al (2015) No longer wearing: investigating
  the abandonment of personal health-tracking technologies on craigslist. In:
  Proceedings of the 2015 ACM international joint conference on pervasive and
  ubiquitous computing, pp 647--658

\bibitem[{for Clinical~Excellence et~al(2007)}]{national2007nice}
for Clinical~Excellence NI, et~al (2007) Nice public health guidance 6
  behaviour change at population, community and individual levels. London, NICE
  (http://www nice org uk/nicemedia/pdf/PH006guidance pdf)

\bibitem[{Consolvo et~al(2006)Consolvo, Everitt, Smith, and
  Landay}]{consolvo2006design}
Consolvo S, Everitt K, Smith I, et~al (2006) Design requirements for
  technologies that encourage physical activity. In: Proceedings of the SIGCHI
  conference on Human Factors in computing systems. ACM, Montréal, Canada, pp
  457--466

\bibitem[{Consolvo et~al(2008)Consolvo, McDonald, Toscos, Chen, Froehlich,
  Harrison, Klasnja, LaMarea, LeGrand, Libby, Smith, and Landay}]{Consolvo2008}
Consolvo S, McDonald DW, Toscos T, et~al (2008) {Activity sensing in the wild:
  A field trial of UbiFit Garden}. In: Conference on Human Factors in Computing
  Systems - Proceedings. ACM, Florence, Italy, pp 1797--1806,
  \doi{10.1145/1357054.1357335}

\bibitem[{Consolvo et~al(2009)Consolvo, McDonald, and
  Landay}]{consolvo2009theory}
Consolvo S, McDonald DW, Landay JA (2009) Theory-driven design strategies for
  technologies that support behavior change in everyday life. In: Proceedings
  of the SIGCHI conference on human factors in computing systems, pp 405--414

\bibitem[{Crawford et~al(1991)Crawford, Jackson, and
  Godbey}]{crawford1991hierarchical}
Crawford DW, Jackson EL, Godbey G (1991) A hierarchical model of leisure
  constraints. Leisure sciences 13(4):309--320

\bibitem[{Cronbach(1951)}]{cronbach1951coefficient}
Cronbach LJ (1951) Coefficient alpha and the internal structure of tests.
  psychometrika 16(3):297--334

\bibitem[{Darnton(2008)}]{darnton2008practical}
Darnton A (2008) Practical guide: An overview of behaviour change models and
  their uses. Government Social Research Unit: www gsr gov
  uk/downloads/resources/behaviour\_change\_review/practical\_guide pdf

\bibitem[{De~Vaus(2002)}]{de2002analyzing}
De~Vaus D (2002) Analyzing social science data: 50 key problems in data
  analysis. sage

\bibitem[{Deci and Ryan(2012)}]{deci2012self}
Deci EL, Ryan RM (2012) Self-determination theory. Handbook of theories of
  social psychology

\bibitem[{Deci and Ryan(2013)}]{deci2013intrinsic}
Deci EL, Ryan RM (2013) Intrinsic motivation and self-determination in human
  behavior. Springer Science \& Business Media

\bibitem[{Duro et~al(2019)Duro, Campos, Rom{\~a}o, and
  Karapanos}]{duro2019motivational}
Duro L, Campos PF, Rom{\~a}o T, et~al (2019) How do motivational text messages
  impact motivation to exercise? implications for the design of activity
  trackers. In: Proceedings of the 13th Biannual Conference of the Italian
  SIGCHI Chapter: Designing the next interaction, pp 1--10

\bibitem[{Epstein et~al(2020)Epstein, Caldeira, Figueiredo, Lu, Silva,
  Williams, Lee, Li, Ahuja, Chen et~al}]{epstein2020mapping}
Epstein DA, Caldeira C, Figueiredo MC, et~al (2020) Mapping and taking stock of
  the personal informatics literature. Proceedings of the ACM on Interactive,
  Mobile, Wearable and Ubiquitous Technologies 4(4):1--38

\bibitem[{Eufic(2014)}]{eufic_2014}
Eufic (2014) Behaviour change models and strategies.
  \urlprefix\url{https://www.eufic.org/en/healthy-living/article/motivating-behaviour-change}

\bibitem[{Filo and Coghlan(2016)}]{filo2016exploring}
Filo K, Coghlan A (2016) Exploring the positive psychology domains of
  well-being activated through charity sport event experiences. Event
  Management 20(2):181--199

\bibitem[{Fogg(2009)}]{fogg2009behavior}
Fogg BJ (2009) A behavior model for persuasive design. In: Proceedings of the
  4th international Conference on Persuasive Technology, pp 1--7

\bibitem[{Gandhi(2015)}]{attrition2}
Gandhi M (2015) Deconstructing the fitbit ipo and s-1.
  \urlprefix\url{https://rockhealth.com/deconstructing-fitbit-s-1/}

\bibitem[{Glanz et~al(2008)Glanz, Rimer, and Viswanath}]{glanz2008health}
Glanz K, Rimer BK, Viswanath K (2008) Health behavior and health education:
  theory, research, and practice. John Wiley \& Sons

\bibitem[{Goldberg(1990)}]{goldberg1990alternative}
Goldberg LR (1990) An alternative" description of personality": the big-five
  factor structure. Journal of personality and social psychology 59(6):1216

\bibitem[{Gouveia et~al(2015)Gouveia, Karapanos, and
  Hassenzahl}]{gouveia2015we}
Gouveia R, Karapanos E, Hassenzahl M (2015) How do we engage with activity
  trackers? a longitudinal study of habito. In: Proceedings of the 2015 ACM
  international joint conference on pervasive and ubiquitous computing, pp
  1305--1316

\bibitem[{Hekler et~al(2013{\natexlab{a}})Hekler, Klasnja, Froehlich, and
  Buman}]{Hekler2013}
Hekler EB, Klasnja P, Froehlich JE, et~al (2013{\natexlab{a}}) {Mind the
  theoretical gap: Interpreting, using, and developing behavioral theory in HCI
  research}. Conference on Human Factors in Computing Systems - Proceedings pp
  3307--3316. \doi{10.1145/2470654.2466452}

\bibitem[{Hekler et~al(2013{\natexlab{b}})Hekler, Klasnja, Froehlich, and
  Buman}]{hekler2013mind}
Hekler EB, Klasnja P, Froehlich JE, et~al (2013{\natexlab{b}}) Mind the
  theoretical gap: interpreting, using, and developing behavioral theory in hci
  research. In: Proceedings of the SIGCHI Conference on Human Factors in
  Computing Systems, pp 3307--3316

\bibitem[{Hinkin(1998)}]{hinkin1998brief}
Hinkin TR (1998) A brief tutorial on the development of measures for use in
  survey questionnaires. Organizational research methods 1(1):104--121

\bibitem[{Jackson and Rucks(1995)}]{jackson1995negotiation}
Jackson EL, Rucks VC (1995) Negotiation of leisure constraints by junior-high
  and high-school students: An exploratory study. Journal of leisure research
  27(1):85--105

\bibitem[{Jackson et~al(1993)Jackson, Crawford, and
  Godbey}]{jackson1993negotiation}
Jackson EL, Crawford DW, Godbey G (1993) Negotiation of leisure constraints.
  Leisure sciences 15(1):1--11

\bibitem[{Josekutty~Thomas et~al(2017)Josekutty~Thomas, Masthoff, and
  Oren}]{josekutty2017personalising}
Josekutty~Thomas R, Masthoff J, Oren N (2017) Personalising healthy eating
  messages to age, gender and personality: using cialdini's principles and
  framing. In: Proceedings of the 22nd International Conference on Intelligent
  User Interfaces Companion, pp 81--84

\bibitem[{Klasnja et~al(2017)Klasnja, Hekler, Korinek, Harlow, and
  Mishra}]{klasnja2017toward}
Klasnja P, Hekler EB, Korinek EV, et~al (2017) Toward usable evidence:
  optimizing knowledge accumulation in hci research on health behavior change.
  In: Proceedings of the 2017 CHI conference on human factors in computing
  systems, pp 3071--3082

\bibitem[{Langer(1975)}]{langer1975illusion}
Langer EJ (1975) The illusion of control. Journal of personality and social
  psychology 32(2):311

\bibitem[{Lee et~al(2011)Lee, Kiesler, and Forlizzi}]{lee2011mining}
Lee MK, Kiesler S, Forlizzi J (2011) Mining behavioral economics to design
  persuasive technology for healthy choices. In: Proceedings of the sigchi
  conference on human factors in computing systems, pp 325--334

\bibitem[{Van~der Lelie(2006)}]{van2006value}
Van~der Lelie C (2006) The value of storyboards in the product design process.
  Personal and ubiquitous computing 10(2-3):159--162

\bibitem[{Leonard(2008)}]{leonard2008richard}
Leonard TC (2008) Richard h. thaler, cass r. sunstein, nudge: Improving
  decisions about health, wealth, and happiness

\bibitem[{Li et~al(2011)Li, Dey, and Forlizzi}]{li2011understanding}
Li I, Dey AK, Forlizzi J (2011) Understanding my data, myself: supporting
  self-reflection with ubicomp technologies. In: Proceedings of the 13th
  international conference on Ubiquitous computing, pp 405--414

\bibitem[{Lin et~al(2006)Lin, Mamykina, Lindtner, Delajoux, and
  Strub}]{lin2006fish}
Lin JJ, Mamykina L, Lindtner S, et~al (2006) Fish’n’steps: Encouraging
  physical activity with an interactive computer game. In: International
  conference on ubiquitous computing, Springer, pp 261--278

\bibitem[{Maddux(1993)}]{maddux1993social}
Maddux JE (1993) Social cognitive models of health and exercise behavior: An
  introduction and review of conceptual issues. Journal of Applied Sport
  Psychology 5(2):116--140

\bibitem[{Mahmud et~al(2019)Mahmud, Fang, Carreiro, Wang, and
  Boyer}]{mahmud2019wearables}
Mahmud MS, Fang H, Carreiro S, et~al (2019) Wearables technology for drug abuse
  detection: A survey of recent advancement. Smart Health 13:100,062

\bibitem[{Mannell and Loucks-Atkinson(2005)}]{mannell2005don}
Mannell R, Loucks-Atkinson A (2005) Why don’t people do what’s “good”
  for them? cross-fertilization among the psychologies of nonparticipation in
  leisure, health, and exercise behaviors. Constraints to leisure 221:232

\bibitem[{Marcus et~al(1992)Marcus, Rossi, Selby, Niaura, and
  Abrams}]{marcus1992stages}
Marcus BH, Rossi JS, Selby VC, et~al (1992) The stages and processes of
  exercise adoption and maintenance in a worksite sample. Health psychology
  11(6):386

\bibitem[{Mejova and Kalimeri(2019)}]{mejova2019effect}
Mejova Y, Kalimeri K (2019) Effect of values and technology use on exercise:
  implications for personalized behavior change interventions. In: Proceedings
  of the 27th ACM Conference on User Modeling, Adaptation and Personalization,
  pp 36--45

\bibitem[{Michie and Johnston(2012)}]{michie2012theories}
Michie S, Johnston M (2012) Theories and techniques of behaviour change:
  Developing a cumulative science of behaviour change

\bibitem[{Michie et~al(2008)Michie, Johnston, Francis, Hardeman, and
  Eccles}]{michie2008theory}
Michie S, Johnston M, Francis J, et~al (2008) From theory to intervention:
  mapping theoretically derived behavioural determinants to behaviour change
  techniques. Applied psychology 57(4):660--680

\bibitem[{Mimiaga et~al(2009)Mimiaga, Reisner, Reilly, Soroudi, and
  Safren}]{MIMIAGA2009203}
Mimiaga MJ, Reisner SL, Reilly L, et~al (2009) Chapter 8 - individual
  interventions. In: Mayer KH, Pizer HF (eds) HIV Prevention. Academic Press,
  San Diego, p 203--239,
  \doi{https://doi.org/10.1016/B978-0-12-374235-3.00008-X},
  \urlprefix\url{https://www.sciencedirect.com/science/article/pii/B978012374235300008X}

\bibitem[{Munson and Consolvo(2012)}]{munson2012exploring}
Munson SA, Consolvo S (2012) Exploring goal-setting, rewards, self-monitoring,
  and sharing to motivate physical activity. In: 2012 6th international
  conference on pervasive computing technologies for healthcare
  (pervasivehealth) and workshops, IEEE, pp 25--32

\bibitem[{Nadirova and Jackson(2000)}]{nadirova2000alternative}
Nadirova A, Jackson EL (2000) Alternative criterion variables against which to
  assess the impacts of constraints to leisure. Journal of leisure research
  32(4):396--405

\bibitem[{Neațu(2015)}]{neatu2015public}
Neațu AM (2015) Public health and behavioral economics. nudging behaviors
  through wearable technology. International Journal of Economic Practices and
  Theories 5(5):518--526

\bibitem[{Niess et~al(2020)Niess, Knaving, Kolb, and
  Wo{\'z}niak}]{niess2020exploring}
Niess J, Knaving K, Kolb A, et~al (2020) Exploring fitness tracker
  visualisations to avoid rumination. In: 22nd International Conference on
  Human-Computer Interaction with Mobile Devices and Services, pp 1--11

\bibitem[{Nunnally(1994)}]{nunnally1994psychometric}
Nunnally JC (1994) Psychometric theory 3E. Tata McGraw-hill education

\bibitem[{Organization(2019)}]{world2019global}
Organization WH (2019) Global action plan on physical activity 2018-2030: more
  active people for a healthier world. World Health Organization, Geneva,
  Switzerland

\bibitem[{Orji and Moffatt(2018)}]{orji2018persuasive}
Orji R, Moffatt K (2018) Persuasive technology for health and wellness:
  State-of-the-art and emerging trends. Health informatics journal 24(1):66--91

\bibitem[{Orji et~al(2017)Orji, Nacke, and Di~Marco}]{orji2017towards}
Orji R, Nacke LE, Di~Marco C (2017) Towards personality-driven persuasive
  health games and gamified systems. In: Proceedings of the 2017 CHI Conference
  on Human Factors in Computing Systems, pp 1015--1027

\bibitem[{Oyebode et~al(2021)Oyebode, Ndulue, Mulchandani, A.~Zamil~Adib,
  Alhasani, and Orji}]{oyebode2021tailoring}
Oyebode O, Ndulue C, Mulchandani D, et~al (2021) Tailoring persuasive and
  behaviour change systems based on stages of change and motivation. In:
  Proceedings of the 2021 CHI Conference on Human Factors in Computing Systems,
  pp 1--19

\bibitem[{Palen et~al(2011)Palen, Caldwell, Smith, Gleeson, and
  Patrick}]{palen2011mixed}
Palen LA, Caldwell LL, Smith EA, et~al (2011) A mixed-method analysis of
  free-time involvement and motivation among adolescents in cape town, south
  africa. Leisure/Loisir 35(3):227--252

\bibitem[{Partners(2014)}]{attrition1}
Partners E (2014) Inside wearables part 1: How behavior change unlocks
  long-term engagement.
  \urlprefix\url{https://medium.com/@endeavourprtnrs/inside-wearable-how-the-science-of-human-behavior-change-offers-the-secret-to-long-term-engagement-a15b3c7d4cf3}

\bibitem[{Peters et~al(2018)Peters, Calvo, and Ryan}]{peters2018designing}
Peters D, Calvo RA, Ryan RM (2018) Designing for motivation, engagement and
  wellbeing in digital experience. Frontiers in psychology 9:797

\bibitem[{Prochaska and Velicer(1997)}]{prochaska1997transtheoretical}
Prochaska JO, Velicer WF (1997) The transtheoretical model of health behavior
  change. American journal of health promotion 12(1):38--48

\bibitem[{Rabbi et~al(2020)Rabbi, Philyaw-Kotov, Li, Li, Rothman, Giragosian,
  Reyes, Gadway, Cunningham, Bonar et~al}]{rabbi2020translating}
Rabbi M, Philyaw-Kotov M, Li J, et~al (2020) Translating behavioral theory into
  technological interventions: Case study of an mhealth app to increase
  self-reporting of substance-use related data. arXiv preprint arXiv:200313545

\bibitem[{Rapp(2019)}]{rapp2019design}
Rapp A (2019) Design fictions for behaviour change: exploring the long-term
  impacts of technology through the creation of fictional future prototypes.
  Behaviour \& Information Technology 38(3):244--272

\bibitem[{Rooksby et~al(2014)Rooksby, Rost, Morrison, and
  Chalmers}]{rooksby2014personal}
Rooksby J, Rost M, Morrison A, et~al (2014) Personal tracking as lived
  informatics. In: Proceedings of the SIGCHI conference on human factors in
  computing systems, pp 1163--1172

\bibitem[{Rotter(1966)}]{rotter1966generalized}
Rotter JB (1966) Generalized expectancies for internal versus external control
  of reinforcement. Psychological monographs: General and applied 80(1):1

\bibitem[{Ryan and Deci(2008)}]{ryan2008self}
Ryan RM, Deci EL (2008) Self-determination theory and the role of basic
  psychological needs in personality and the organization of behavior. The
  Guilford Press

\bibitem[{Stevens et~al(2019)Stevens, Rees, and Polman}]{stevens2019social}
Stevens M, Rees T, Polman R (2019) Social identification, exercise
  participation, and positive exercise experiences: Evidence from parkrun.
  Journal of Sports Sciences 37(2):221--228

\bibitem[{Strack and Deutsch(2004)}]{strack2004reflective}
Strack F, Deutsch R (2004) Reflective and impulsive determinants of social
  behavior. Personality and social psychology review 8(3):220--247

\bibitem[{Suh et~al(2016)Suh, Shahriaree, Hekler, and
  Kientz}]{suh2016developing}
Suh H, Shahriaree N, Hekler EB, et~al (2016) Developing and validating the user
  burden scale: A tool for assessing user burden in computing systems. In:
  Proceedings of the 2016 CHI conference on human factors in computing systems,
  pp 3988--3999

\bibitem[{Tang and Kay(2017)}]{tang2017harnessing}
Tang LM, Kay J (2017) Harnessing long term physical activity data—how
  long-term trackers use data and how an adherence-based interface supports new
  insights. Proceedings of the ACM on Interactive, Mobile, Wearable and
  Ubiquitous Technologies 1(2):1--28

\bibitem[{Todorov et~al(2002)Todorov, Chaiken, and
  Henderson}]{todorov2002heuristic}
Todorov A, Chaiken S, Henderson MD (2002) The heuristic-systematic model of
  social information processing. The persuasion handbook: Developments in
  theory and practice pp 195--211

\bibitem[{Truong et~al(2006)Truong, Hayes, and Abowd}]{10.1145/1142405.1142410}
Truong KN, Hayes GR, Abowd GD (2006) Storyboarding: An empirical determination
  of best practices and effective guidelines. In: Proceedings of the 6th
  Conference on Designing Interactive Systems. Association for Computing
  Machinery, New York, NY, USA, DIS '06, p 12–21,
  \doi{10.1145/1142405.1142410},
  \urlprefix\url{https://doi.org/10.1145/1142405.1142410}

\bibitem[{Tversky and Kahneman(1974)}]{tversky1974judgment}
Tversky A, Kahneman D (1974) Judgment under uncertainty: Heuristics and biases.
  science 185(4157):1124--1131

\bibitem[{Vailshery(2021)}]{vailshery_2021}
Vailshery LS (2021) Wearables sales worldwide by region 2015-2022.
  \urlprefix\url{https://www.statista.com/statistics/490231/wearable-devices-worldwide-by-region/}

\bibitem[{Weissinger et~al(1992)Weissinger, Caldwell, and
  Bandalos}]{weissinger1992relation}
Weissinger E, Caldwell LL, Bandalos DL (1992) Relation between intrinsic
  motivation and boredom in leisure time. Leisure Sciences 14(4):317--325

\bibitem[{White(1959)}]{white1959motivation}
White RW (1959) Motivation reconsidered: the concept of competence.
  Psychological review 66(5):297

\bibitem[{Wikipedia(2021)}]{wikipedia_2021}
Wikipedia (2021) Storyboard.
  \urlprefix\url{https://en.wikipedia.org/wiki/Storyboard}

\bibitem[{Wilde et~al(2018)Wilde, Ward, Sewell, M{\"u}ller, and
  Wark}]{wilde2018apps}
Wilde LJ, Ward G, Sewell L, et~al (2018) Apps and wearables for monitoring
  physical activity and sedentary behaviour: A qualitative systematic review
  protocol on barriers and facilitators. Digital health 4:2055207618776,454

\bibitem[{Yfantidou et~al(2021)Yfantidou, Sermpezis, and
  Vakali}]{yfantidou2021self}
Yfantidou S, Sermpezis P, Vakali A (2021) Self-tracking technology for mhealth:
  A systematic review and the past self framework. arXiv preprint
  arXiv:210411483

\end{thebibliography}

\end{document}